\documentclass[11pt,letter,thmsb]{article}

\usepackage[letterpaper, margin=1in]{geometry}
\usepackage[toc,page,title,titletoc,header]{appendix}
\usepackage{fancyhdr}
\usepackage{caption}
\usepackage{subcaption}
\usepackage{graphicx}
\usepackage[T1]{fontenc}
\usepackage{units}
\usepackage{amsmath}
\usepackage{mathtools}
\usepackage{multirow}
\usepackage{array}
\usepackage{enumitem}
\usepackage{amssymb}
\usepackage{mathrsfs}
\usepackage{bm}
\usepackage{bbm}
\usepackage[margin=1cm]{caption}
\usepackage{pgfplots}
\usepackage{color, soul}
\usepackage{booktabs}
\usepackage{hyperref}
\usepackage{appendix}
\usepackage[labelfont=bf]{caption}
\usepackage{threeparttable}
\usepackage{dcolumn}
\newcolumntype{d}[1]{D..{#1}} 

\DeclareMathOperator{\E}{E}
 
\DeclareMathOperator*{\argmin}{arg\,min}
\renewcommand*\Pr{\mathop{\rm Pr}\nolimits} 

\let\originalleft\left
\let\originalright\right
\renewcommand{\left}{\mathopen{}\mathclose\bgroup\originalleft}
\renewcommand{\right}{\aftergroup\egroup\originalright}

\begin{document}


\title{
	Machine Predictions and Human Decisions with Variation in Payoffs and Skill%
	\thanks{We benefited from very helpful feedback by Jason Abaluck, David Chan, Tomaso Duso, Daniel Ershov, Mogens Fosgerau, Matthew Gentzkow, Ulrich Kaiser, Jonathan Kolstad, Chuck Manski, Jeanine Mikl\'os-Thal, Ye\c{s}im Orhun, Bertel Schjerning, Stephan Seiler, Jann Spiess, Christoph Wolf, and participants at the 13th Digital Economics Conference in Toulouse, the 6th International Conference on Computational Social Science virtually at MIT, the Econometric Society World Congress 2020 virtually at Bocconi, the Electronic Health Economics Colloquium, as well as in seminars at DIW Berlin, University of Copenhagen, and University Paris-Sud. We are deeply indebted to Lars Bjerrum and Gloria Cristina Cordoba Currea for providing their expertise on diagnostics and antibiotic prescribing in Danish primary care as well as to Jenny Dahl Knudsen, Sidsel Kyst, and Rolf Magnus Arpi for enabling us to work with the microbiological laboratory data. Financial support from the European Research Council (ERC) under the European Union's Horizon 2020 research and innovation programme (grant agreement no. 802450) is gratefully acknowledged.}

}

\author{
	Michael Allan Ribers%
		\thanks{DIW Berlin; Department of Economics, University of Copenhagen; and BCCP. michael.ribers@econ.ku.dk}
	\and
	Hannes Ullrich%
		\thanks{DIW Berlin; Department of Economics, University of Copenhagen; BCCP; and CESifo. hullrich@diw.de.}
}

\date{November 2020}

\maketitle

\vspace{-1cm}

\thispagestyle{empty} \setcounter{page}{0}

\begin{abstract}

\noindent 
Human decision-making differs due to variation in both incentives and available information. This constitutes a substantial challenge for the evaluation of whether and how machine learning predictions can improve decision outcomes. We propose a framework that incorporates machine learning on large-scale data into a choice model featuring heterogeneity in decision maker payoff functions and predictive skill. We apply this framework to the major health policy problem of improving the efficiency in antibiotic prescribing in primary care, one of the leading causes of antibiotic resistance. Our analysis reveals large variation in physicians' skill to diagnose bacterial infections and in how physicians trade off the externality inherent in antibiotic use against its curative benefit. Counterfactual policy simulations show that the combination of machine learning predictions with physician diagnostic skill results in a 25.4 percent reduction in prescribing and achieves the largest welfare gains compared to alternative policies for both estimated physician as well as conservative social planner preference weights on the antibiotic resistance externality.
\end{abstract} 

\vspace{-0.2cm}


\newpage

\section{Introduction}

Machine learning methods and the increasing availability of high-quality, large-scale data provide new opportunities to design welfare improving policies for a broad set of problems with prediction at their core (Kleinberg et al. 2015; Agrawal, Gans, and Goldfarb 2018; Athey 2018). Prominent examples include bail decisions in criminal justice, hiring, detecting social service fraud, healthcare provision, and labor market assistance programs. In numerous situations, machine learning can provide a standardized, data-based risk assessment. Yet, evaluating the potential of machine learning predictions relative to the status quo is complicated by the fact that human decisions are outcomes of individual incentives and prediction technologies. Importantly, observed heterogeneity in decisions can be a result of variation in both (Chan, Gentzkow, and Yu 2019). In addition, in settings studied so far, the outcome of interest is difficult to observe and often sampled selectively as a result of human decisions.\footnote{For example, in health settings such as the diagnosis of heart attacks considered in Mullainathan and Obermeyer (2019), a patient is defined as recovered when no subsequent return to the hospital is observed. In Kleinberg et al. (2018), the machine learning algorithm can only be trained on observed recidivism by defendants to whom judges decided to grant bail, the very decision to which machine learning predictions are being compared.} These empirical challenges make it difficult to learn about whether and by which mechanisms policies using machine learning predictions can improve outcomes.

In this paper, we consider a prime example of a policy challenge where prediction is key, the emerging health crisis due to increasing antibiotic resistance. Antibiotic use is considered the main driver of antibiotic resistance and inefficient prescribing can be decreased by accurate diagnosis of bacterial vs.~other causes of infections (WHO 2014).\footnote{Antibiotics are used to treat bacterial infections by killing or inhibiting growth of bacteria in the body. Their effectiveness is decreasing due to antibiotic resistant bacteria threatening to render simple infections, such as pneumonia or infections in wounds, a fatal risk. In the US alone, antibiotic-resistant infections result in an estimated 23,000 deaths, \$20 billion in direct healthcare costs, and \$35 billion in lost productivity each year (CDC 2013).} Specifically, we study antibiotic treatment decisions for urinary tract infections (UTI) in primary care, combining rich, administrative data on individual patients with diagnostic outcomes from gold standard microbiological laboratory test results in Denmark.\footnote{UTI are one of the most common classes of bacterial infections. Foxman (2002) reports almost half of all women contract a UTI once in their lifetime. In the US, yearly UTI-related healthcare costs including workplace absences are estimated at \$3.5 billion (Flores-Mireles et al. 2015), with 10 percent of all women receiving antibiotic treatment for UTI (Bjerrum and Lind\ae k 2015). Primary care accounts for 90 percent of prescriptions in Europe and for 75 percent of prescriptions in Denmark (Llor and Bjerrum 2014; Danish Ministry of Health 2017).} These test outcomes are observed by physicians but with a delay of several days, often corresponding to a complete course of antibiotic treatment. Due to the acute nature of UTI, physicians must make a treatment decision before test results arrive. Therefore, we observe outcomes independently of the human decision of interest, a feature which we exploit to study the mechanisms leading to antibiotic prescription decisions. The example we consider here is important because delayed diagnostic results with simultaneous urgency to treat are a common challenge in health care; for example, in biopsies for malignant tumors or testing for tuberculosis. Understanding the role of instant diagnostic information for treatment decisions is central to improving such decisions (Cassidy and Manski 2019).

We combine machine learning predictions with a binary choice model governing physicians' treatment decisions. The model accommodates two main steps physicians take when treating patients. First, the underlying cause of reported symptoms must be assessed. Second, given their assessment, physicians decide whether or not to prescribe an antibiotic, the standard treatment for bacterial infections. The risk assessment of a bacterial cause of infection in the first step depends on a physician's diagnostic skill. We introduce machine learning predictions in the model by decomposing physicians' diagnostic skill in two dimensions. The first is the analysis of information encoded in observable data amenable to machine learning methods. For example, physicians may spend time observe some of a patient's personal characteristics and medical history which can be informative about the probability of having a bacterial UTI. The second dimension is the acquisition and interpretation of diagnostic information in clinical practice which is not commonly available to policy makers. For example, patients describe symptoms and their general health condition. Physicians may also perform rapid diagnostic technologies, notably urine dipstick and microscopic analysis (Davenport et al. 2017). While the dipstick analysis is a standard procedure, using microscopy requires additional equipment and specific training. This separation of physicians' diagnostic skill by observable and unobservable diagnostic information provides a method to evaluate the impact of combining machine learning predictions with physician-specific expertise on treatment decisions. 

In addition to the diagnostic problem, antibiotic prescribing involves an important trade off. Antibiotics are effective treatments for bacterial infections but using them causes a negative externality by promoting increased antibiotic resistance. For every treatment decision, physicians must consider this trade-off and weigh patients' sickness cost against the social cost of antibiotic resistance. Formulating the model allows to distinguish whether diagnostic skill or preferences drive decisions. Therefore, the combination of predictions with a model of treatment choice provides a framework for the evaluation of policies using machine learning predictions as well as policies targeting preferences. Potential policies can include the provision of personalized predictions or targeting physicians' payoffs, for example via behavioral nudging or using taxes.

For the identification of diagnostic skill and physician payoff functions, we follow results in Chan, Gentzkow, and Yu (2019). They rely on an assumption of quasi-random assignment of suspected pneumonia cases to radiologists evaluating chest X-rays because the true outcomes are only observed for patients for whom medical treatment has not yet begun. The fact that in our setting physicians make an antibiotic treatment decision before test results are known allows us to predict patient types using rich observable data and to use these predictions to condition on expected test outcomes. The crucial condition for this strategy is that expected test outcomes can be predicted without bias. We present statistical test results that cannot reject unbiasedness of predictions for nearly all physicians. In addition, we provide evidence that unobservable information obtained from point of care rapid tests is unlikely to drive patient selection into laboratory testing. We show that by observing test outcomes independently of treatment decisions, conditioning on individual predicted test outcomes identifies physician skill and preference parameters in the structural model.

To predict diagnostic test results, we build on Ribers and Ullrich (2020) who train an extreme gradient boosting machine learning algorithm on high-dimensional, administrative data from Denmark. The outcome is an indicator variable taking the value of one when bacteria are isolated in patient urine samples submitted for microbiological laboratory testing. The covariates in the prediction model include a rich set of patient-level information, such as gender, age, detailed employment status and type, education, income, civil status and more, past antibiotic prescriptions, past microbiological test results, medical outpatient claims histories, hospitalization records, as well as the same information on each individuals' household members. Machine learning applied to these data predicts out of sample realizations of bacterial UTI well. We document large heterogeneity in physician decisions evaluated by true and false positive rates as well as by the degree of predictive information of their decisions. The predictive information of physicians' prescription decisions is positively associated with the propensity to send test samples to the microbiological laboratory and with the share of female physicians in a primary care clinic.

We estimate three parameters of the structural model for each primary care clinic using the machine learning predictions of patient types. The first two measure the accuracy of diagnostic information physicians use, while the third parameter governs the trade-off physicians solve in making treatment decisions. The mean estimated signal noise parameter on patient-type information is larger than signal noise on clinical diagnostic information, implying that, on average, physicians rely more on information from in-clinic diagnostics than on observing patient types. We document significant heterogeneity in the estimated noise parameters. Over one-third of physicians make no use of patient type information encoded in observable data. The mean estimated weight of the social cost of antibiotic resistance relative to an individual patient's sickness cost is 0.56 with a standard deviation of the distribution across physicians of 0.13. This implies that the mean physician weighs the social cost of increasing antibiotic resistance due to one antibiotic prescription slightly above one half the health benefit of curing one patient. 

To gain insights about potential reasons for physician heterogeneity, we correlate the parameter estimates with primary care clinic characteristics. We find that clinics with more patients per physician use patient type information less. The noise parameter on clinical diagnostic information is positively correlated with mean physician age and negatively associated with the intensity of laboratory testing, suggesting that physicians with higher skill may be younger on average and rely more on high-quality diagnostic technologies.

We use the structural model to evaluate three counterfactual policies. The first provides physicians with the machine learning prediction for each patient. Here, we hold physicians' payoff functions and clinical diagnostic information fixed. We find that this policy decreases overall prescribing by 25.4 percent (4,146 prescriptions) and overprescribing, prescriptions to patients without bacterial infections, by 44.4 percent (2,831 prescriptions). The number of treated bacterial infections decreases by 13.2 percent (1,315 prescriptions). The second counterfactual policy changes physicians' preference parameters while holding patient type and clinical diagnostic information fixed. We increase the payoff parameter such that the reduction in overall prescribing is equivalent to the reduction achieved by providing predictions to physicians. This can be achieved by policies such as nudging or an antibiotic tax shifting the weight on the antibiotic resistance externality. The policy reduces overprescribing by 33.4 percent (2,133 prescriptions). Yet, without improving diagnostic information this policy induces adverse effects: the number of treated bacterial infections decreases by 20.0 percent (1,990 prescriptions), significantly more than in the first counterfactual. These results illustrate the usefulness of separating the prediction and decision step in the structural model. The effects of interventions attempting to incentivize behavior according to social objectives can be considered independently from interventions aimed at purely improving diagnostic information.

We also consider a third counterfactual which is not based on the structural model. It differs from the first two policies because it ignores physician incentives and substitutes rather than complements diagnostic skill. The policy we evaluate considers a corner case, which does not require knowledge of physicians' payoff functions. To achieve this, it distributes prescriptions from low-risk to high-risk patients, holding the number of treated bacterial infections fixed. Such a policy effectively replaces physician decisions for risk ranges in which the machine learning predictions are most accurate. Much of the existing work evaluating machine learning predictions for policy follows comparable approaches (Bayati et al. 2014; Chalfin et al. 2016; Kleinberg et al. 2018; Yelin et al. 2019; Ribers and Ullrich 2019, Hastings, Howison, and Inman 2020). This counterfactual policy reduces prescribing by 10.0 percent (1,633 prescriptions) without treating fewer patients suffering from a bacterial infection, and reduces overprescribing by 25.2 percent (1,607 prescriptions). 

Computing gains in payoffs for the full range of preference parameters, we find the best policy depends on a social planner's weight on the antibiotic resistance externality. The third counterfactual guarantees positive aggregate payoff gains for all weights. Yet, the counterfactuals which give physicians full discretion and make use of their skill generate the largest payoff gains for sufficiently high weights on the externality, at the mean estimated physician-level preference weights and above.

We contribute to the existing literature in several ways. First, we identify a high-stakes policy problem in which combining signal from a machine learning algorithm and human expertise can play an important role for decision making. The existing literature has considered a variety of settings, in with machine learning can deliver useful predictions. However, evaluating prediction tools in the field is often difficult for ethical, legal, or practical reasons. Therefore, for considering the use of machine learning in practice, it appears particularly useful to evaluate potential interventions \emph{ex ante}, taking into account the information decision makers have absent an intervention. The existing literature has stopped short of measuring heterogenous skill and preferences driving decisions that machine learning predictions aim to improve. Kang et al. (2013) show that online reviews can help predict restaurants' sanitary conditions for hygiene inspections but do not compare these predictions to authorities' decisions which restaurants to inspect. Andini et al. (2018) predict household consumption responses for the targeting of a tax rebate program in Italy but consider a decision rule based exclusively on machine learning predictions. 

Several existing studies consider health care applications but without incorporating human expertise in the evaluation of machine learning predictions. Hastings, Howison, and Inman (2019) predict the riskiness of opioid prescriptions and impose constraints on decisions based on predictions. Yelin et al. (2019) predict molecule-specific antibiotic resistance probabilities, conditional on knowing which bacteria are present, and improve prescription efficiency by redistributing molecules while holding the distribution of prescribed molecules fixed. Ribers and Ullrich (2019) consider policies which redistribute antibiotic prescriptions from patients with low to high predicted risk of having a bacterial infection. They find improvements are possible by partly overriding physicians' decisions. Yet, in many situations physicians have diagnostic information that machine learning predictions cannot provide. 

Kleinberg et al. (2018) make an important contribution by using machine learning to evaluate the potential improvements of judges' bail decisions, where only crimes committed by released defendants can be observed. Their framework requires the assumption that judges' risk prediction skill is homogenous and preferences, reflected in judge leniency, are the source of variation in decisions. Currie and MacLeod (2017) allow for heterogeneity in skill but assume homogenous preferences to evaluate the counterfactual of reassigning C-sections from low- to high-risk pregnancies. While they focus on the effects of improving surgical and decision-making skill, we focus on the potential of data-driven predictions to complement physician skill. Importantly, we allow for heterogeneity in skill and preferences. This flexibility is crucial for combining machine learning predicted risk with physician expertise when heterogenous preferences are likely an important driver of decision-making.

Finally, a large literature explores policies to tackle the problem of growing antibiotic resistance. We contribute to this work by providing evidence for the potential effectiveness of prediction-based demand side policies aimed at curbing antibiotic resistance. Urinary tract infections are the second most common bacterial infection accounting for a bulk of human antibiotic consumption. The literature considering policy interventions includes Laxminarayan et al. (2013) on prescription surveillance and stewardship programs, Bennett, Hung, and Lauderdale (2015) on general practitioner competition in Taiwan, Currie, Lin, and Meng (2014) and Das et al. (2016) on financial incentives for physicians in China and India, Kwon and Jun (2015) on peer effects in Korea, and Hallsworth et al. (2016) on communication of social norms by alerting high-prescribers of their status in the UK. While this research has identified various ways to reduce antibiotic prescribing, the cost in terms of potential undertreatment are often difficult to measure. Without improving diagnostic information, many policies may lead to unintended consequences and costs. Making the distinction between diagnostic information and preferences is pivotal for designing and evaluating efficient policy measures. We provide a framework to make this distinction explicit. 

The remainder of the paper is organized as follows. Section 2 presents the institutional background and data. Section 3 shows the machine learning prediction results and Section 4 inspects observed heterogeneity in prescription decisions. Section 5 develops the structural model of physician prescription choice when skill and preferences vary, discusses identification and estimation, and estimation results. Section 6 presents counterfactual policy evaluations and Section 7 concludes.

\section{Danish administrative data and laboratory test results}

We use Danish administrative registry data which cover a vast array of information including patient and patient household members' detailed socioeconomic data as well as antibiotic prescription histories, general practice insurance claims and hospitalization records. Notably, the coherent use of unique personal identifiers enables us to merge registries as well as connect individuals to laboratory test results acquired from two major Danish hospitals. 

\subsection{The Danish healthcare system}
Denmark has a universal and tax financed single payer health care system with general practitioners as the primary gatekeepers. Every person living in Denmark is allocated to a general practitioner by a list-system within a fixed geographic radius around the home address. Patients can switch physicians from their initial assignment at a small cost but most stick with their assigned general practitioner. Although primary care clinics operate as privately owned businesses, all fees for services are collectively negotiated between the national union of general practitioners and the public health insurer. Importantly, physicians do not generate earnings by prescribing drugs to patients who have to purchase their prescriptions from local pharmacies. In 2012, Denmark had 2,200 primary care clinics with a median size of just above one general practitioner per clinic (M\o ller Pederson, Sahl Andersen, and S\o ndergaard 2012). Throughout the paper, we will use physician and clinic interchangeably because most of our medical transaction data are observed at the clinic level. General practitioners are responsible for prescribing approximately 75 percent of the human consumed systemic antibiotics in Denmark (Danish Ministry of Health 2017).

\subsection{Analysis sample based on clinical microbiological laboratory test results}
Individual-level clinical microbiological laboratory test results comprise the central data set of our analysis. Particularly, it contains the outcome we aim to predict, a binary outcome indicating if bacteria were isolated when a urine sample was acquired from a patient consulting a general practice physician. We acquired clinical microbiological laboratory test results from Herlev hospital and Hvidovre hospital, the two major hospitals in Denmark's capital region covering a catchment area of roughly 1.7 million inhabitants, nearly one third of the Danish population, for the period of January 2010 to December 2012. The laboratory data provides the bacterial species and relevant antibiotic resistances when bacteria are detected in a patient sample. In addition, patient and clinic identifiers and information on the biological sample type, the test acquisition date, sample arrival date at the laboratory, and test response date is provided. A total of 2,579,617 microbiological samples are observed in the time period with submissions from both general practitioner clinics and hospitals. Urine samples constitute 477,609 samples out of which 156,694 are submitted by general practitioners, the focus of our application. Bacteria were isolated in approximately one out of three urine samples, both overall and among the general practitioner submitted samples. We further restrict the number of observations in order to focus on consultations that constitute a first contact with a physician within the patient's treatment spell. Hence, we exclude test observations where the patient received a systemic antibiotic prescription or were previously tested within 28 days prior to the sample acquisition date. Lastly, we also exclude pregnant women from our analysis as both the test decision, including many mandatory tests during pregnancy, and the prescription decision cannot be compared to the typical non-emergency patient. 

The set of test observations used for machine learning is comprised of 74,511 test results for urine samples taken during initial consultations with men or non-pregnant women in a non-emergency setting consulting physicians in 688 primary care clinics. For this set, nearly all of the test procedures lasted two or more days during which general practitioners must decide under uncertainty. Since we know the precise timing of urine sample acquisitions and the test response date, we can determine exactly whether physicians' prescribe antibiotics with or without knowledge of the test result. By focussing on consultations during which physicians submitted a urine sample for microbiological laboratory testing, we ensure that test outcomes are observed for all patients regardless of the physicians' prescription decisions. We so avoid the selective labels problem tackled by Kleinberg et al. (2018), a common challenge when evaluating counterfactuals in machine learning applications, and advance what can be learnt about the role of machine learning predicted risk for treatment decisions conditional on laboratory testing. The drawback of this approach is a lack of generalizability of our results to prescription occasions that did not include patient microbiological testing. However, laboratory testing for bacterial UTI is common and indicated in clinical practice when point-of-care diagnostics are inconclusive (Davenport et al. 2017). When a physician decides to test, the value of diagnostic information is presumably high so that the prediction-based policies proposed here improve upon situations in which physicians are making decisions under significant uncertainty.

\subsection{Danish national registries}
The administrative data provided by Statistics Denmark covers the entire population of Denmark between January 1, 2002, and December 31, 2012. The registries can be linked for all individuals in our laboratory data. For each, we observe a comprehensive set of socioeconomic and demographic variables, the complete prescription history of systemic antibiotics (\emph{L\ae gemiddeldatabasen}), hospitalizations (\emph{Landspatientregisteret}), and general practitioner insurance claims (\emph{Sygesikringsregisteret}).

The demographic data include gender, age, education, occupation, income, marriage and family status, home municipality, immigration status and place of origin, and lastly includes household member identifiers which allows us to identify the patients' family members and add their demographic data as well as the laboratory data and the data from the following registries.
The data on systemic antibiotic prescriptions contain slightly more than 35 million purchased prescriptions. We observe the date of purchase, patient and prescribing primary care clinic identifiers, anatomical therapeutic chemical drug classification, drug name, price, indication of use, purchased package size and defined daily dose. It should be noted that the indication of use is imprecise in the sense that many prescriptions are given with a UTI indication but prescriptions for UTI are also given with a more generic indication, e.g. against infection, or without indication at all.
The hospitalization data comprise all patient contacts with hospitals, including ambulatory visits. The data contain observations on hospitalizations of more than 2 million unique individuals per year over since 2002 and includes information on hospitalization admittance and discharge dates, procedures performed, type of hospitalization (ambulatory, emergency, etc), primary and secondary diagnoses and the number of total bed days.
Lastly, the insurance claims data cover all Danish general practitioner clinic services provided to the Danish population of patients. The claims data are comprised of approximately 100 million claims per year and include physician and patient identifiers, the week of consultation, and services used. Among other, the claims data allow us to identify pregnant women from mandatory pregnancy-associated examinations who we exclude from the analysis.
The combination of the laboratory data and the administrative registers yields a vector $x_{it}$ of 1,215 predictor variables for patient $i$ at time $t$. The predictor variables can be grouped into categories including patient characteristics and test timing, patient past prescriptions, patient past laboratory test results, patient past hospitalizations, patient past general practice insurance claims, household members' past prescriptions, household members' past laboratory test results, household members' past hospitalizations, household members' past hospitalizations, and household members' past general practice insurance claims. All contained historic data relative to the test acquisition time are, in principle, observable to the physician at the time of the consultation.

\section{Machine learning predictions}

We use prediction results from Ribers and Ullrich (2020) who train an extreme gradient boosting algorithm, proposed by Friedman, Hastie, and Tibshirani (2000) and Friedman (2001), to predict if patients suffer from bacterial UTI based on information contained in laboratory tests and a rich set of individual patient data.\footnote{We use the extreme gradient boosting algorithm implemented in xgboost for R.} We create 24 monthly out-of-sample evaluation partitions from January 2011 to December 2012, and use all data prior to the respective test partition as training data. In summarizing the prediction results and the subsequent empirical analysis, we additionally drop clinics with fewer than 100 test observations to provide a sufficient level of statistical power. The resulting sample has 42,480 observations distributed across 194 primary care clinics. 

\begin{figure*}[h!]
\centering
\includegraphics[height=8cm]{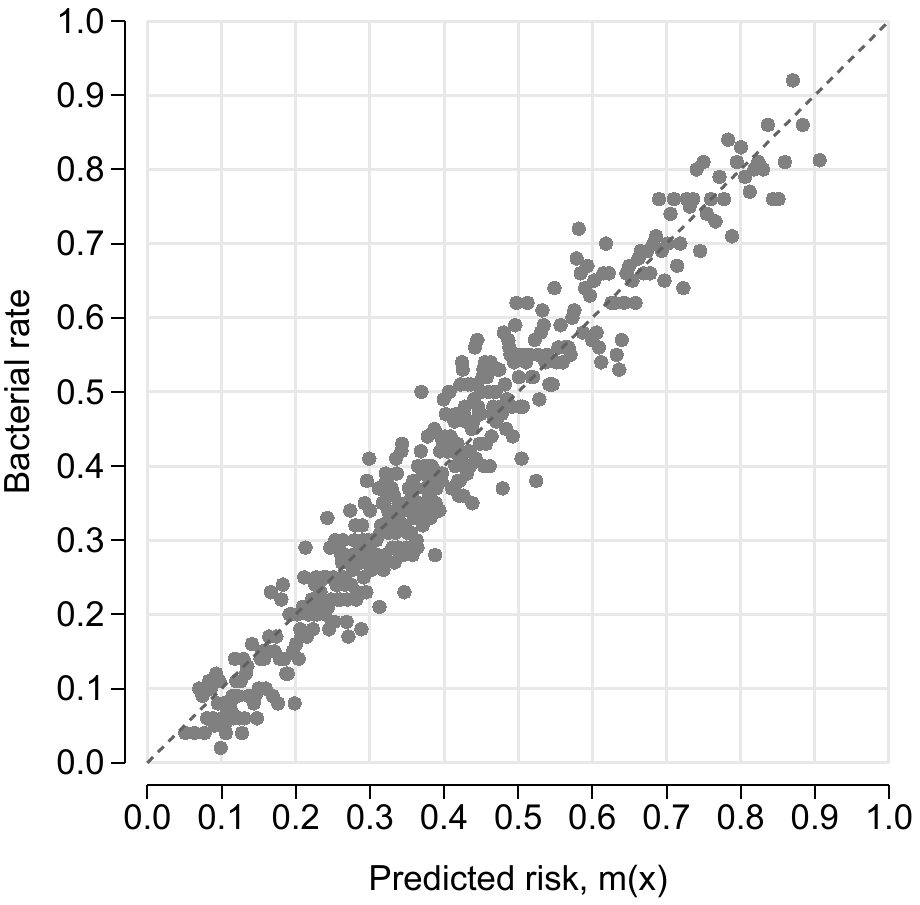}
\caption{Laboratory test results relative to machine predictions of bacterial test outcomes. Spheres represent averages over 100 tested patients sorted by predicted risk.}
\label{fig:bact_vs_predict}
\end{figure*}
We illustrate the quality of the machine learning predictions $m(x)$ of the binary outcome $y$, indicating the result of a microbiological test result, in Figure \ref{fig:bact_vs_predict} which plots the average test results against the average out of sample predicted risk. Every sphere represents a bin containing 100 patients where patients are assigned to bins sorted by their predicted risk. Outcomes are close to the 45 degree line throughout the risk distribution, showing that the algorithm on average correctly predicts bacterial risk. A common measure of prediction quality for binary outcomes is the area under the receiver operating curve (AUC) for out-of-sample observations. Our prediction function for positive bacterial test outcomes has an AUC equal to 0.728. In comparison, Kleinberg et al. (2018) report a comparable AUC of 0.707 in predicting defendants' failure to appear in court in criminal justice.

\section{Heterogeneity in prescription decisions}
\label{sec:predict_decide}

\begin{figure*}[h!]
	\centering
		\includegraphics[height=8cm]{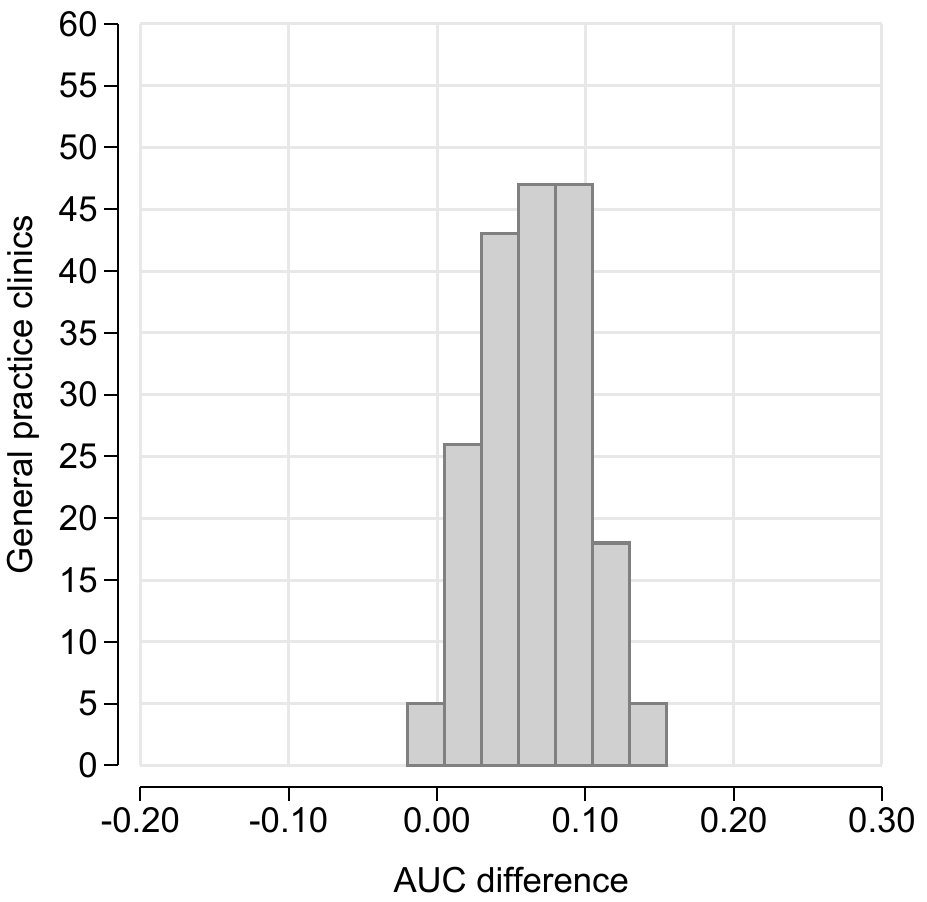}
		\caption{Distribution of physician-specific changes in AUC due to treatment choice as a predictor. Bins include at least five observations to ensure anonymity.}
	\label{fig:auc_diff}
\end{figure*}

To inspect potential heterogeneity in physician decisions, we first consider the predictive value of physician decisions for laboratory test outcomes by including prescription choices in the set of predictors. Figure \ref{fig:auc_diff} shows the clinic-specific difference in AUC between two prediction models. One includes the physician's treatment choice as a potential predictor and one excludes it. Including physician choice increases the AUC on average and for most clinics. The observed heterogeneity suggests that combining physician information and data-based predictions at the physician level may have potential.

To learn about potential correlates with this measure, we link these results to a set of clinic characteristics which we observe for 117 out of the total of 194 clinics. Table \ref{tab:gp_aucreg} in Appendix \ref{app:gp_aucreg} shows the coefficients of a linear regression of the change in AUC on clinic characteristics. The number of laboratory tests ordered per patient are positively associated with improvement in prediction due to information contained in treatment choices. One interpretation of this observation is that physicians with more frequent exposure to patients with urinary tract infections are better at identifying bacterial infection causes. Being a female physician is also positively correlated with the amount of predictive information in prescription decisions while physician age is negatively associated. While we can give no causal interpretation to the parameters estimated in this analysis, the correlations hint towards differences in the use of diagnostic technologies across clinics based on types of patients as well as physician age and gender. It is important to note that the interpretation of the change in AUC due to the treatment choice predictor as physician skill is confounded by the fact that physician choice is the outcome of an optimization problem according to the physician's objective function. A physician may know the true bacterial outcome nearly perfectly for each patient but still decide to give a prescription irrespective of beliefs. If so, the predictive information measured here does not, without further assumptions, reflect the diagnostic information physicians hold.

Because these observations are limited in the extent to which they can inform policies, we borrow intuition from Chan, Gentzkow, and Yu (2019) who view the diagnostic problem through the lens of a standard binary classification problem. A common tool to study such a problem is the receiver operating characteristic (ROC) curve, which is the set of all trade-offs between false positive and true positive rates a given classification technology allows. For antibiotic prescribing, a false positive is considered an overprescription, that is a prescription to a person who did not suffer from a bacterial infection, and a true positive is an effective prescription of an antibiotic to a person with a bacterial infection. At one extreme of this set every patient with a bacterial infection can be given an antibiotic, at the cost of complete overprescribing. Conversely, overprescribing can be completely avoided at the cost of giving no antibiotics to any patients, including ones with a bacterial infection. 

The achievable trade offs between these extremes depend on a physicians' skill to diagnose whether an illness is caused by a bacterial infection or not. Given this skill, the position on the associated ROC curve reflects the physician's choice of trade off between false and true positive rates. We can directly calculate physicians' false and true positives rates and plot their location in the ROC space because we observe the disease state for every tested patient irrespective of prescription decisions. This is due to the fact that the only conclusive way to diagnose a bacterial urinary tract infection is a microbiological analysis, which returns results only after several days. In the meantime, physicians must make prescription decisions under uncertainty, relying on their diagnostic skill absent a laboratory test result. By observing physicians' locations we can learn about the diagnostic information available to physicians at the time of prescription decisions.

\begin{figure}[h!]
    	\centering
    	\includegraphics[height=8cm]{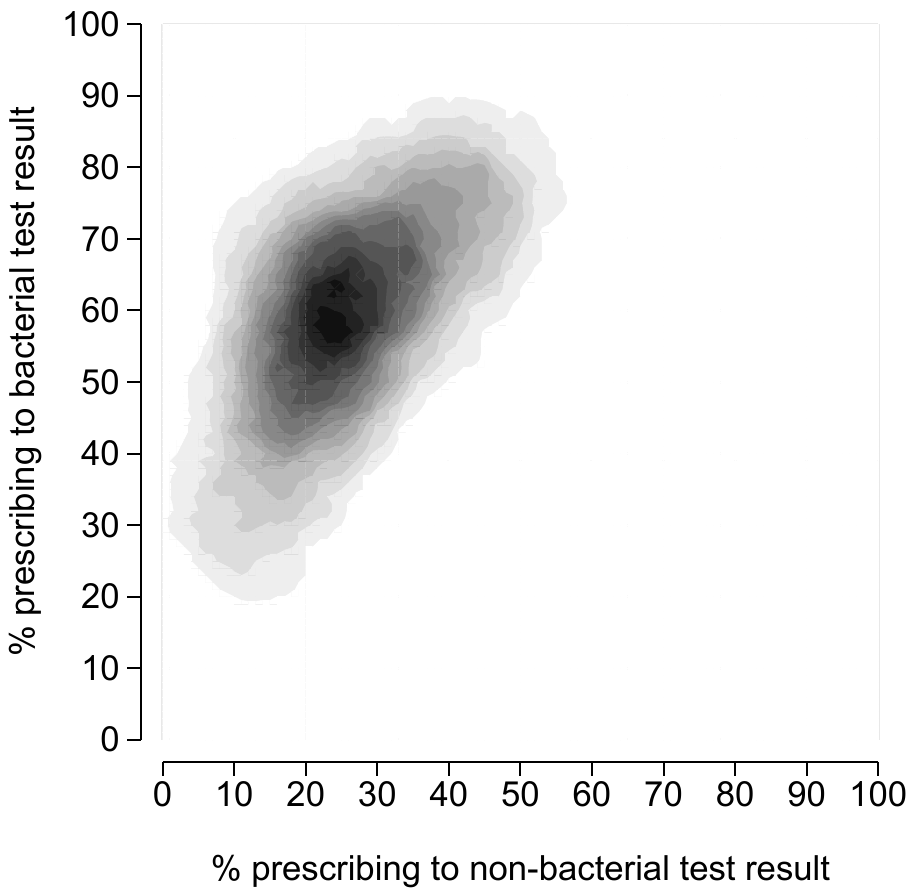}
	\caption{Heat map of physicians' true and false positive rates before test results. For anonymity, aggregated into areas of five or more physicians.}
	\label{fig:prescr_bact_contour}
\end{figure}

Figure \ref{fig:prescr_bact_contour} shows a heat map of prescription rates relative to negative and positive bacterial test outcomes. Physician prescribing relative to bacterial outcomes varies widely. Physicians' location close to the origin place is suggestive of a large weight on the antibiotic resistance externality relative to individual sickness cost, hence low levels of overprescribing but also low levels of appropriate prescribing. Physicians to the top right are more intense prescribers which suggests a low weight on the antibiotic resistance externality relative to individual sickness cost. The plot suggests that general practitioners in Denmark do remarkably well in avoiding prescribing to non-bacterial cases while at the same time prescribing to a high share of bacterial infections. Yet, the significant variation both away from as well as parallel to the diagonal line suggest that policies may be able to improve decision outcomes by enhancing diagnostic prediction as well as incentivizing physicians to choose different trade offs.\\

\section{A treatment choice model with variation in payoffs and skill}
\label{sec:model}

We propose a formal framework that combines machine learning predictions with a model of primary care providers' treatment choice allowing for heterogenous payoff functions and skill. The model follows Chan, Gentzkow, and Yu (2019) by separating the individual physicians' treatment choice problem from the preceding step of forming predictions. Specifically, we consider physician skill in two dimensions: diagnosis based on observable background information, as also used by the machine learning algorithm, and diagnosis based on unobservable clinical information available only to the physician. The distinction of these two types of diagnostic skill and the payoff function in a model of physician prescription choice provides a systematic framework to analyze the effects of counterfactual policies that improve diagnostic skill or manipulate physicians' payoff functions.

\subsubsection*{Sickness realization}

We model patient $i$'s sickness realization as determined by a latent index, $\nu_{i}$, such that the patient has a bacterially caused infection according to
\begin{equation}
	y_{i} = \mathbbm{1} [\nu_{i} > \bar{\nu}],
\label{eq:sick-def}
\end{equation}
where $\bar{\nu}$ is a common threshold across all patients. The latent patient index is normally distributed with mean $\tau_{i}$, the patient's type, such that
\begin{equation}
	\nu_{i} \sim \mathcal{N}(\tau_{i}, \sigma_{\nu}^{2}).
\label{eq:sick-latent}
\end{equation}
We do not require any assumptions on the distribution of patient types, $\tau_{i}$, across physicians. Instead, we recover $\tau_{i}$ from $m(x_{i}) = \E\{ y_{i} \mid x_{i} \} $, that is, by assigning patient type as the machine learning predicted risk conditional on observables $x_{i}$.

\subsubsection*{Prediction}

In clinical practice, when patients present symptoms of a urinary tract infection, physicians gather information about patients' true sickness state by observing $x_{i}$, including $i$'s personal characteristics and medical histories. Some physicians may research patients' medical histories in more detail than others so that risk assessment based on observable data depends on analytical skill. We assume that the physician receives a noisy signal about patient $i$'s type, where lower noise implies higher skill,
\begin{equation}
	 \xi_{ij} \sim \mathcal{N}(\tau_{i}, \sigma_{\xi_{j}}^{2}).
\label{eq:xi-latent}
\end{equation}
In addition, physicians can acquire clinical diagnostic information by observing patients' health condition and performing either one or both of the rapid diagnostic technologies available today: urine dipstick and microscopic analysis (Davenport et al. 2017). The dipstick analysis is standard procedure but microscopic analysis requires additional equipment and specific training. Errors in interpreting dipstick results and performing microscopic analysis may introduce substantial variation in diagnostic skill in this setting, an observation that has been documented in medical decision making more generally (Hoffrage et al. 2000, Pallin et al. 2014). Thus, physicians receive a noisy signal by observing clinical diagnostic information,
\begin{equation}
	 \eta_{ij} \sim \mathcal{N}(\nu_{i}, \sigma_{\eta_{j}}^{2}).
\label{eq:eta-latent}
\end{equation} 
Figure \ref{fig:model_signals} in Appendix \ref{app:model_signals} illustrates the distribution of $\nu_i$ and the signals $\xi_{ij}$ and $\eta_{ij}$ for an example patient and physician.

Given both patient type and clinical diagnostic signals, the physician forms her posterior beliefs, $v_{ij}$, about the latent patient index according to
\begin{equation}
	\nu_{ij} \mid \xi_{ij}, \eta_{ij} 
			\sim \mathcal{N} (\mu_{ij},\sigma_{ij}^{2})
\end{equation}
where the posterior mean and variance are given by
\begin{equation}
	\mu_{ij} =  \frac{\xi_{ij} \sigma_{\eta_{j}}^{2} + \eta_{ij} (\sigma_{\xi_{j}}^{2}+\sigma_{\nu}^{2}) }{\sigma_{\xi_{j}}^{2} + \sigma_{\nu}^{2} + \sigma_{\eta_{j}}^{2}}
	\qquad \text{and} \qquad
	\sigma_{j}^{2} = \frac{(\sigma_{\xi_{j}}^{2}+\sigma_{\nu}^{2})\sigma_{\eta_{j}}^{2}}{\sigma_{\xi_{j}}^{2} + \sigma_{\nu}^{2} + \sigma_{\eta_{j}}^{2}}.
\label{eq:posterior_m_s}
\end{equation}

\subsubsection*{Treatment choice}

Physician $j$'s payoff function reflects the trade off between a patient suffering sickness cost $a_j$ from delaying prescribing until a test result is available, and the social cost of prescribing $\beta_j$ associated with a potential increase in antibiotic resistance due to antibiotic use. While the social cost is incurred for every antibiotic prescribed, the sickness cost of waiting is only incurred by untreated patients suffering from a bacterial infection. Antibiotic treatment is only curative and alleviates sickness if a patient suffers from a bacterially caused infection. The payoff function at a patient's initial consultation can therefore be written as
\begin{equation}
		\pi(\, d, y \, ; \alpha_{j}, \beta_{j})  = -\alpha_j y(1-d)-\beta_j d ,
\label{eqn_policy_payoff_function}
\end{equation}
where $d$ is an indicator for the decision to prescribe an antibiotic. We assume $0<\beta_j<\alpha_j$ such that prescribing is always optimal when an infection is known to be bacterial with certainty.

A payoff-maximizing physician then proceeds to prescribe an antibiotic if and only if the expected sickness cost is larger than the social cost of prescribing:
\begin{align}
\begin{split}
	\beta_{j} < \alpha_{j} \left(1-\Phi \left( \frac{\bar{\nu}-\mu_{ij}}{\sigma_{ij}} \right) \right)
\quad \Leftrightarrow \quad
	\mu_{ij} \: > \:  v_{j}^{*}(\sigma_{\xi_{j}},\sigma_{\eta_{j}},\alpha_{j},\beta_{j})
\end{split}
\end{align}
where
\begin{equation}
\label{eqn:cutoff}
	v_{j}^{*}(\sigma_{\xi_{j}},\sigma_{\eta_{j}},\alpha_{j},\beta_{j}) = \bar{v} - \sigma_{j} \Phi^{-1} \left( 1 - \frac{\beta_{j}}{\alpha_{j}} \right) .
\end{equation}
As patient types are only identified relative to their distance from the sickness threshold, $\bar{v}$, in terms of units of $\sigma_{\nu}$, we set $\bar{v}=0$ and $\sigma_{\nu} = 1$ resulting in the final prescription rule:
\begin{equation}
	d_{ij} \mid \xi_{ij}, \eta_{ij} 
		= \mathbbm{1} [\; 0 < \underbracket{ \xi_{ij} \sigma_{\eta_{j}}^{2} + \eta_{ij} (\sigma_{\xi_{j}}^{2}+1) + \sqrt{(1+\sigma_{\xi_{j}}^{2}+\sigma_{\eta_{j}}^{2})(\sigma_{\xi_{j}}^{2}+1)\sigma_{\eta_{j}}^{2}} \Phi^{-1}(1-\frac{\beta_j}{\alpha_j}) }_{ g(\xi_{ij},\eta_{ij} \mid \alpha_{j}, \beta_{j}, \sigma_{\xi_{j}}, \sigma_{\eta_{j}}) } \;] .
\end{equation}
We assume physicians hold correct beliefs about their own skill. Low signal variance reflects high skill. The comparative statics with respect to $\sigma_{\xi_{j}}$, $\sigma_{\eta_{j}}$, and $\beta_{j}/\alpha_{j}$ are intuitive. The larger a physician's weight on the antibiotic resistance externality relative to individual patients' sickness cost, the less likely she is to prescribe an antibiotic. The effect of the two skill parameters $\sigma_{\xi_{j}}$ and $\sigma_{\eta_{j}}$ is ambiguous. Low skill, reflected in large parameter values, increases the last term of the inequality and hence the probability to prescribe. On the other hand, with low skill, large signal realizations can turn the sign of the first or second terms and lead to prescriptions for non-bacterial cases or vice versa.

\subsection{Identification} 

Our identification strategy follows the arguments in Chan, Gentzkow, and Yu (2019) who show that, under the assumption of random assignment, the skill and preference parameters of their model are identified. Their setting requires a design using random assignment because they only observe the true outcome for a selected part of their sample. We exploit a feature which is common for many medical diagnostic and treatment situations. Consequential treatment decisions must often be made before complete diagnostic test results are available. In our data, the waiting period for test results was typically at least two days and often three to five days, exceeding the duration of an antibiotic treatment. The \emph{ex post} observation of true test outcomes, regardless of physician prescription decisions at the time of test acquisition, allows us to predict patient types using rich observable data and to use these predictions to condition on expected test outcomes.

Equations (\ref{eq:sick-latent}) to (\ref{eq:posterior_m_s}) combined with a physician's observed patient types, $m(x)$, determine the shape of all potential ROC curves as a function of physician skill parameters $\sigma_{\xi_{j}}$ and $\sigma_{\eta_{j}}$. The ROC curves are guaranteed to be smooth and no two ROC curves can intersect. This ensures that each physician, defined by her observed false positive and false negative rates, lies on only one ROC curve. As the model includes two signals, $\xi$ and $\eta$, ROC curves may not always be generated from a unique combination of $\sigma_{\xi_{j}}$ and $\sigma_{\eta_{j}}$. The two physician skill parameters are then separately identified by observations for which the patient type and clinical diagnostic signal contradict and the physician's decision reflects which of the signals she relies on. In our application, identification of $\sigma_{\xi_{j}}$ is achieved by observing treatment decisions for patients with similar type signals $\xi$ and variation in clinical diagnostic signals $\eta$, and identification of $\sigma_{\eta_{j}}$ analogously by observing treatment decisions for patients with similar $\eta$ and variation in $\xi$.

Having pinned down the physician's ROC curve, the preference parameter $\beta_{j}$ is then identified by the cutoff value in equation (\ref{eqn:cutoff}) which moves the estimated physician along the ROC curve from the bottom left, $\beta=1$, to the top right, $\beta=0$. Simulating ROC curves based on estimated $\sigma_{\xi_{j}}$, $\sigma_{\eta_{j}}$, and $\beta_j$, we find that in our data each physician depicted in Figure \ref{fig:prescr_bact_contour} lies on the ROC corresponding to their skill and preference parameters.\footnote{For reasons of anonymity we cannot show individual physician data points.}

The crucial condition for this strategy is that the expected test outcome can be predicted without bias. Bias can, for example, be introduced if physicians rely on unobservables to select whom to test.\footnote{Bias may in principle also be introduced by patients' selection of which physician to consult. In Denmark, primary care providers are assigned by an individual's municipality of residence. Switching away from these default assignments is possible but uncommon. Therefore, physicians treating UTI are almost completely determined by location of residence. The data we use for prediction contain information about patients' location of residence in addition to the described socioeconomic and health data, allowing for the prediction algorithm to use this information.} We cannot test directly whether the machine learning predictions recover patient types without bias because we only observe one single patient sickness realization. Instead, we test whether the physician-level sum of the sickness indicator concurs with the set of patient type predictions. If predicted patient types are true, the physician-level sum of sickness realizations follows the Poisson-Binomial distribution, that is, the sum of non-identical independent Bernoulli trials. Using this test, we cannot reject machine learning patient type predictions for 160 out of 194 physicians, that is 83 percent, at the five percent level. We verify that the estimation results are robust to leaving out physicians for whom we reject that patient type predictions are true on average at the five percent level. 

\begin{figure*}[h!]
	\centering
		\includegraphics[height=8cm]{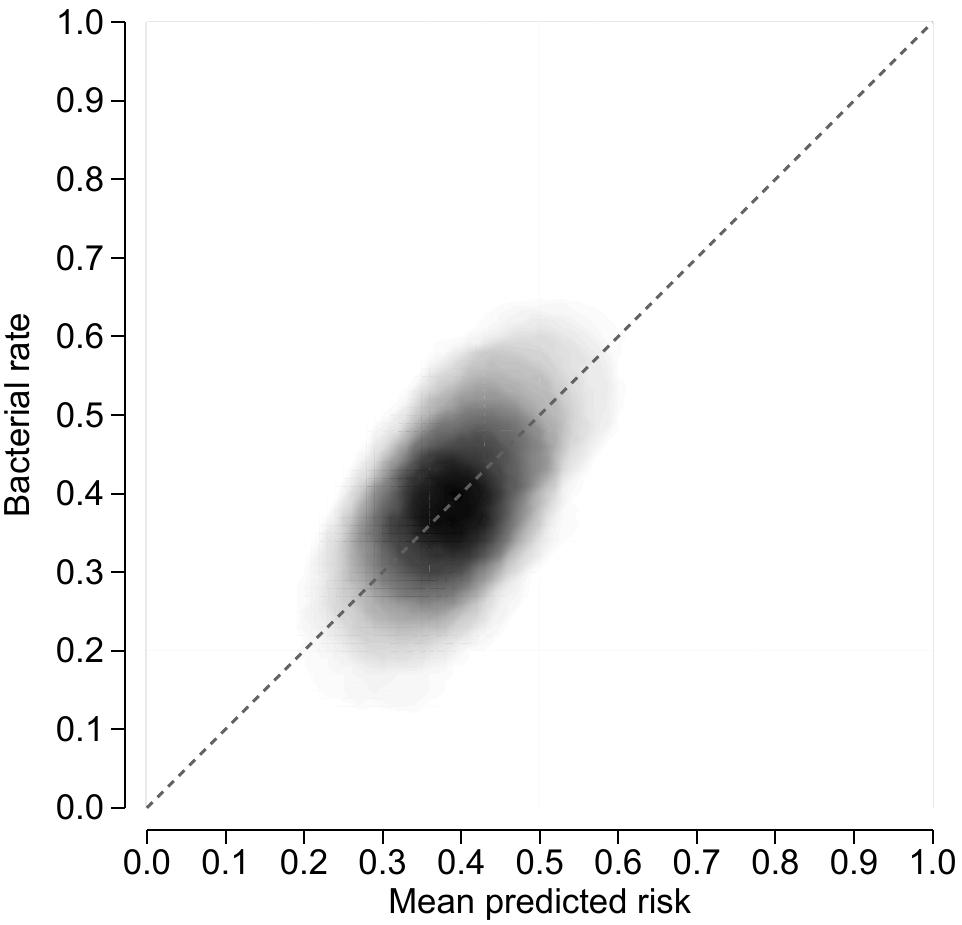}
		\caption{Heatmap of physician mean bacterial rate and predicted bacterial risk. Areas with five physicians or more are plotted.}
	\label{fig:uncrate_gp}
\end{figure*}

Figure \ref{fig:uncrate_gp} shows physician-level mean bacterial rates and mean predicted risk are both strongly centered around the value of 0.4. The remaining variation around the center of the sphere follows the diagonal suggesting further that risk predictions are unbiased at the physician-level. If physicians selected patients to be tested based on unobservables that cannot be captured by a machine learning algorithm, we would expect to see variation off the diagonal.

To consider the potential role of unobservable information for physicians' decisions who to test further, we also inspect the balance of the types of bacteria found in tests as well as their antibiotic resistances. Different bacteria have varying difficulty of detection by in-clinic diagnostics such as dipstick and microscopic analysis. Further, they can lead to different symptoms and disease severities. These may be important sources of information contained in unobservables, which we assume do not drive the decision to procure a microbiological test. Abaluck et al. (2016) show that variation in skill should drive the bacterial rate of tests, that is test yields, and not the absolute number of tests. In clinical practice, microbiological analysis in the laboratory is typically required if the urine dipstick analysis is inconclusive regarding the bacterial cause of an infection (Davenport et al. 2017). Point of care tests can give an indication for the presence of some but not all types of bacteria. If skill varies regarding the signal physicians can observe from such point of care tests or symptom assessment, and this affects the decision to obtain a laboratory diagnostic, this variation would be reflected in differing rates of different bacteria across physicians with different overall test yield. For example, if there is variation in the ability to identify E.coli bacteria from point of care tests, then we would see variation in the share of positive E.coli test results across physicians with varying test yield.
\begin{table}[h!]
\centering
\small
\begin{threeparttable}
\caption{Balance of types of bacterial infection causes}
\label{tab:bactbalance_medianresid}
\begin{tabular}{ @{} l @{\extracolsep{24mm}} c @{\extracolsep{12mm}} c @{\extracolsep{12mm}} c @{} } \\[-1cm]
\toprule\\[-0.8cm]
& $\underline{E_j[y]}$ & $\overline{E_j[y]}$ & $\Delta$ \\
\midrule\\[-0.8cm]
E.coli		& 0.67& 0.68 &0.014		\\[-0.9em] &{\scriptsize(0.08)}&{\scriptsize (0.06)}&{\scriptsize(0.010)}\\[-0.5em]
E.faecalis	 	& 0.05& 0.04 &-0.014	\\[-0.9em] &{\scriptsize(0.04)}&{\scriptsize(0.02)}& {\scriptsize(0.005)}\\[-0.5em]
Enterococcus	& 0.03& 0.04 &0.007		\\[-0.9em] &{\scriptsize(0.03)}&{\scriptsize(0.03)}& {\scriptsize(0.004)}\\[-0.5em]
K.pneumoniae	& 0.04& 0.05 &0.007		\\[-0.9em] &{\scriptsize(0.03)}&{\scriptsize(0.02)}& {\scriptsize(0.004)}\\[-0.5em]
S. agalactiae	& 0.05& 0.04 &-0.009	\\[-0.9em] &{\scriptsize(0.04)}&{\scriptsize(0.02)}& {\scriptsize(0.005)}\\[-0.5em]
Others		& 0.16& 0.16 &0.005		\\[-0.9em] &{\scriptsize(0.05)}&{\scriptsize(0.04)}& {\scriptsize(0.007)}\\[-0.3em]
\midrule\\[-0.8cm]  
J01CA01		& 0.40& 0.39 &-0.006	\\[-0.9em] &{\scriptsize(0.08)}&{\scriptsize(0.06)}& {\scriptsize(0.010)}\\[-0.5em]
J01CA04 		& 0.02& 0.02 & 0.004	\\[-0.9em] &{\scriptsize(0.02)}&{\scriptsize(0.01)}& {\scriptsize(0.002)}\\[-0.5em]
J01CA11		& 0.26& 0.23 &-0.026	\\[-0.9em] &{\scriptsize(0.07)}&{\scriptsize(0.05)}& {\scriptsize(0.008)}\\[-0.5em]
J01DC02		& 0.09& 0.08 &-0.007	\\[-0.9em] &{\scriptsize(0.04)}&{\scriptsize(0.04)}& {\scriptsize(0.006)}\\[-0.5em]
J01DD13		& 0.03& 0.03 & 0.002	\\[-0.9em] &{\scriptsize(0.02)}&{\scriptsize(0.02)}& {\scriptsize(0.003)}\\[-0.5em]
J01EA01		& 0.22& 0.20 &-0.014	\\[-0.9em] &{\scriptsize(0.07)}&{\scriptsize(0.05)}& {\scriptsize(0.008)}\\[-0.5em]
J01EB02		& 0.35& 0.33 &-0.019	\\[-0.9em] &{\scriptsize(0.08)}&{\scriptsize(0.06)}& {\scriptsize(0.010)}\\[-0.5em]
J01MA02		& 0.10& 0.09 &-0.004	\\[-0.9em] &{\scriptsize(0.05)}&{\scriptsize(0.03)}& {\scriptsize(0.006)}\\[-0.5em]
J01MB02		& 0.10& 0.09 &-0.003	\\[-0.9em] &{\scriptsize(0.05)}&{\scriptsize(0.03)}& {\scriptsize(0.006)}\\[-0.5em]
J01XE01		& 0.07& 0.07 & 0.002	\\[-0.9em] &{\scriptsize(0.04)}&{\scriptsize(0.04)}& {\scriptsize(0.006)}\\[-0.3em]
\midrule\\[-0.8cm]
Number of cases 	&20447&22033&\\[-0.3em]
Number of clinics	&97&97&\\[-0.3em]
\bottomrule
\end{tabular}
\vspace{-0.2cm}
\footnotesize Notes: This table reports mean bacterial species and resistance rates for physicians above and below the median of mean bacterial rates $E_j[y]$. Physician-level means and standard deviations are weighted by physician-level numbers of observations. 
\end{threeparttable}
\end{table}

To investigate this, we split physicians into two groups, those above and below the median clinic-level mean bacterial rate. Table \ref{tab:bactbalance_medianresid} reports the descriptive statistics of physician-level shares of bacteria species for these two groups in positive test results, where we weigh the physician-level mean and standard deviations by their numbers of observations. The small differences in size and the structural similarity across bacteria suggests physicians do not systematically select which patients to test based on informative signals from point of care tests. Likewise, physicians may vary in their knowledge about the prevalence of antibiotic resistance for specific patients or in the population. Such knowledge may also influence the decision to make use of laboratory diagnostics. Table \ref{tab:bactbalance_medianresid} shows that little variation in resistance rates can be observed between physicians with low and high test yield. The balance across nearly all bacteria and antibiotic molecules provides evidence that information at the point of care, which is unobservable to us, is not a strong driver of the selection of patients receiving a laboratory tested.

\subsection{Estimation}
We estimate the model by simulated maximum likelihood using observed data on prescription decisions, $d_{it}$, sickness realizations, $y_{it}$, and patient types $\tau_{i}$ recovered from random forest predictions $m(x_{i})$. We normalize $\alpha_j=1$ because only the ratio $\beta_{j}/\alpha_{j}$ is identified.
The simulated likelihood contribution from a single observation follows from
\begin{equation} \label{eq:llh_i}
	\mathcal{L}_{ij} (d_{ij} \mid \Theta_{j},  y_{i}, m(x_{i})) = \left\{
		\begin{tabular}{l l}
			$\Pr \left\{ g(\xi_{ij},\eta_{ij} \mid \beta_{j}, \sigma_{\xi_{j}},\sigma_{\eta_{ij}})>0 \mid \Theta_{j}, y_{i}, m(x_{i}) \right\}$	& if $d_{ij}=1$ \\
			$\Pr \left\{ g(\xi_{ij},\eta_{ij} \mid \beta_{j}, \sigma_{\xi_{j}}, \sigma_{\eta_{ij}})<0, \mid \Theta_{j}, y_{i}, m(x_{i}) \right\}$	& if $d_{ij}=0$.
		\end{tabular} \right. 
\end{equation}
where $\Theta = \{ \beta_{j}, \sigma_{\xi_{j}}, \sigma_{\eta_{j}} \}$ and $\xi_{ij}$ and $\eta_{ij}$ are simulated conditional on observed $y_{i}$ using the distributional assumptions in equations (\ref{eq:sick-latent}), (\ref{eq:xi-latent}) and (\ref{eq:eta-latent}). Appendix \ref{app:sim_smooth} explains the procedure to simulate the probabilities in equation \ref{eq:llh_i}.
Defining $\mathcal{I}_{j}$ as the set of patients consulting physican $j$, the joint likelihood over outcomes $\bm{d}_{j}=\{ d_{ij} \}_{i\in \mathcal{I}_{j}}$ is given by
\begin{equation}
	\mathcal{L}_{j}(\bm{d}_{j} \mid \Theta_{j}, \bm{y}_{j}, m(\bm{x}_{j})) = \prod_{i\in \mathcal{I}_{j}} \mathcal{L}_{ij} (d_{ij} \mid \Theta_{j}, y_{i}, m(x_{i})).
\end{equation}
where $\bm{y}_{j} = \{ y_{i} \}_{i\in \mathcal{I}_{j}}$ and $m(\bm{x}_{j}) = \{ m(x_{i}) \}_{i\in \mathcal{I}_{j}}$. Physician skill and preferences can now be recovered for physician $j$ from
\begin{equation}
	\hat{\Theta}_{j} = \argmin_{\Theta_{j}} \sum_{i\in \mathcal{I}_{j}} \log \mathcal{L}_{ij}(d_{ij} \mid \Theta_{j}, y_{i}, m(x_{i})) .
\end{equation}
Estimating $\hat{\Theta}_{j}$ for every physician allows us to recover the nonparametric physician heterogeneity distribution.\footnote{To estimate the parameters we maximize the simulated likelihood using 1,000 modified latin hypercube sampling draws proposed by Hess, Train, and Polak (2006) and a quasi-Newton method with analytical gradients provided in Appendix \ref{agradients}. Using the Nelder-Mead method or simulated annealing, a global optimization algorithm, yields nearly identical results.}

\subsection{Estimation results}

Table \ref{tab:parest} reports the means and standard deviations of $\hat{\Theta}_{j}$. The means of the noise parameters for patient type ($\sigma_{\xi_{j}}$) and clinical diagnostic information ($\sigma_{\eta_{j}}$) are large, 6.38 and 2.18. Interestingly, mean $\sigma_{\xi_{j}}$ is markedly larger than $\sigma_{\eta_{j}}$, meaning that physicians rely more on clinical diagnostic information than on information obtained from observing patient types. This result suggests that providing patient type information in the form of machine learning predicted risk should improve physicians' ability to predict the bacterial cause of infections. The extent to which patient type and clinical diagnostic information is used in decisions varies significantly between clinics, as reflected in the standard deviations of the estimates of $\sigma_{\xi_{j}}$ and $\sigma_{\eta_{j}}$. The mean value of 0.56 of the preference parameter estimates, bounded by 0 and 1 by the assumption that $0<\beta_j<\alpha_j$, suggests conservative physicians on average. The mean physician weighs the social cost of increasing antibiotic resistance due to one antibiotic prescription slightly above one half the health benefit of curing one patient. The standard deviation of 0.13 reflects substantial heterogeneity in how physicians solve this trade-off. 
\begin{table}[h!]
\centering
\begin{threeparttable}
\caption{Distribution of parameter estimates}
\label{tab:parest}
\begin{tabular}{ @{} l @{\extracolsep{20mm}} c @{\extracolsep{3mm}} c @{} } \\[-1cm]
\toprule
									& Mean & (SD)  	\\
\midrule
Type signal noise, 			$\sigma_{\xi_{j}}$ 	& 6.38& (3.59) \\
Diagnostic signal noise, 		$\sigma_{\eta_{j}}$ 	& 2.18& (1.40) \\
Payoff function parameter, 	$\beta_j$ 			& 0.56& (0.13) \\
\bottomrule
\end{tabular}
\vspace{-0.2cm}
\footnotesize Notes: This table reports the means and standard deviations of the distribution of parameter estimates over 194 physician clinics. The model is estimated separately for each clinic.
\end{threeparttable}
\end{table}

Figures \ref{fig:est_struct_par_xieta} to \ref{fig:est_struct_par_bxi} in Appendix \ref{sec:est_struct_par} show the distributions of parameter estimates. For anonymization we show heatmaps and do not report values in areas containing less than five clinics. The distribution of the clinical diagnostic skill parameter $\sigma_{\eta_{j}}$ is concentrated in the area between 1 and 3. The noise parameter $\sigma_{\xi_{j}}$ measuring the extent to which physicians make use of patient type information is more dispersed between 1 and 7. The more pronounced concentration of $\sigma_{\eta_{j}}$ estimates suggests that the majority of physicians makes use of clinical diagnostic information even if significant heterogeneity remains. The large estimate of $\sigma_{\xi_{j}}$ on average suggests that providing machine learning predictions can improve physician information significantly. In particular, we find a relevant number of physicians with very large $\sigma_{\xi_{j}}$ estimates. In Figure \ref{fig:est_struct_par_xieta}, physicians with estimated $\sigma_{\xi_{j}}>6$ account for 40\% of all physicians. This group does not appear to use patient type information encoded in observable data. Therefore, combining systematic information in predictions $m(x_i)$ with valuable clinical diagnostic information used by these physicians may substantially improve decisions. Figures \ref{fig:est_struct_par_beta} and \ref{fig:est_struct_par_bxi} do not show a systematic relationship between the estimated payoff weights and both noise parameters. Figures \ref{fig:bs_struct_par_xi} to \ref{fig:bs_struct_par_b} in Appendix \ref{sec:bs_struct_par} show the 95\% confidence intervals for physician-level parameter estimates, computed by bootstrapping at the physician-level. Sorting the estimates by their value, we see that the variance of skill parameter estimates increases in the size of the estimates. The estimates of $\sigma_{\eta_{j}}$ have tight confidence intervals at lower values and throughout from below. In simulations illustrated in Figure \ref{fig:sim_rel_sdxi} in Appendix \ref{sec:sim_rel_sdxi} we show for relevant parameter ranges that, with increasing size of $\sigma_\xi$, variation in the parameter has a vanishingly small effect on physician decisions. This also explains the observation that the lower bound of the confidence intervals follows the parameter estimates more tightly than the upper bound. The confidence intervals for payoff function parameter estimates is uniform over the full range.

We also investigate how well the model fits the data. Figure \ref{fig:modelfit} in Appendix \ref{sec:est_struct_fit} shows the distributions of the observed mean, over-, and underprescribing rates and their simulated counterparts based on the parameter estimates. The simulated distributions closely resemble the observed data.

\subsection{Observed heterogeneity}

To investigate potential sources of heterogeneity across primary care clinics, we correlate parameter estimates with observable clinic characteristics. We aggregate individual physician characteristics to the primary care clinic level because prescriptions are observed for clinics. Due to data limitations we are able to merge characteristics for a subset of 117 out of the total of 194 clinics. Linear regression results of the parameters estimates on clinic characteristics in Table \ref{tab:par_het} show several interesting patterns. 
\begin{table}[h!]
\centering
\small
\begin{threeparttable}
\caption{Parameter estimates and physician characteristics}
\label{tab:par_het}
\begin{tabular}{ @{} l @{\extracolsep{10mm}} r @{\extracolsep{3mm}} l @{\extracolsep{7mm}} r @{\extracolsep{3mm}} l @{\extracolsep{7mm}} r @{\extracolsep{3mm}} l @{} } \\[-1cm]
\toprule
						 & \multicolumn{6}{c}{Linear regression}  \\
\cline{2-7}
						 & \multicolumn{2}{c}{Type signal noise}  & \multicolumn{2}{c}{Clinical signal noise}  & \multicolumn{2}{c}{Preferences}  \\[-0.5em]
$N = 117$					 & \multicolumn{2}{c}{$\hat{\sigma}_{\xi_{j}}$}  & \multicolumn{2}{c}{$\hat{\sigma}_{\eta_{j}}$}  & \multicolumn{2}{c}{$\hat{\beta}_j$}  \\
\midrule
Patients per physician 		& 0.02& [-1.67,\ 1.72] 	& -0.28 & [-0.66,\ 0.11]	&  0.02 &	[-0.05,\ 0.09]\\
Laboratory tests per patient  	&-0.67& [-1.90,\ 0.55]    	& -0.52 & [-0.94, -0.11]	& -0.02 &	[-0.06,\ 0.03]\\
Mean number of physicians  	&-0.44& [-1.86,\ 0.98]    	&  0.07 & [-0.31,\ 0.45]	& -0.03 &	[-0.08,\ 0.03]\\
Mean age of physicians  		&-0.03& [-4.09,\ 4.02]    	&  2.30 & [\ 0.50,\ 4.10]	&  0.09 &	[-0.08,\ 0.27]\\
Share of female physicians  	&-0.15& [-1.03,\ 0.72]    	&-0.001 & [-0.31,\ 0.31]	&  0.003 &	[-0.03,\ 0.04]\\
\midrule
R$^2$  					&\multicolumn{2}{c}{0.02}  & \multicolumn{2}{c}{0.12} & \multicolumn{2}{c}{0.03}	\\
\bottomrule
\end{tabular}
\vspace{-0.2cm}
\footnotesize Notes: This table presents coefficients for three linear regressions where the outcomes are the three parameter estimates of the physician choice model summarized in Table \ref{tab:parest} and the correlates are clinic-level physician characteristics. Mean values are used for multi-physician clinics. Heteroskedasticity-robust 95\% confidence intervals are reported in brackets.
\end{threeparttable}
\end{table}
For the estimate of noise in physicians' use of information encoded in observables $x_i$, $\hat{\sigma}_{\xi_{j}}$, all confidence intervals of covariate coefficients cover the zero. Inspecting the coefficients, less noise is associated with female physicians and higher propensity to request laboratory diagnostic for urine sample, which may reflect more extensive experience treating patients with urinary tract infection. Primary care clinics with more physicians are associated with lower noise estimates. The noise parameter for clinical diagnostic information, $\hat{\sigma}_{\eta_{j}}$, is negatively correlated with the number of laboratory tests a clinic requested per patient. The confidence interval on this correlation excludes the zero. Clinics with higher test intensity may have higher skill in deciding to do additional clinical tests or the ability to better extract information from test diagnostics. Noise in clinical diagnostic information is positively associated with the mean age of physicians in a clinic, with the confidence interval excluding the zero. Several narratives would support this correlation. For example, older physicians might rely more on their clinical experience and personal knowledge of recurring patients than on costly diagnostic tests. Alternatively, they may be less likely to purchase new diagnostic equipment and rely on existing tools they are accustomed to. The results for preference parameter estimates, $\hat{\beta}_j$, are more difficult to interpret because only the ratio between weights $\beta_j$ and $\alpha_j$ is identified. Clinics with on average older physicians seem to have larger weights on the antibiotic resistance externality relative to the weight on alleviating patients' sickness cost, while this ratio is negatively associated with the number of laboratory tests per patient and clinic size measured as the number of physicians.

\section{Counterfactual policy evaluation}

We evaluate counterfactual policies by comparing their induced decision outcomes $\delta$ with the observed physician prescriptions $\delta^0$. Changes in overall prescribing are defined as $\Delta \delta$, changes in overprescribing, i.e. antibiotic prescriptions given to patients without a bacterial infection, as $\Delta \delta\neg y$, and changes in the number of correctly treated bacterial UTI patients as $\Delta \delta y$.

Table \ref{tab:cf_results} shows policy outcomes for three counterfactual interventions.\footnote{We report nearly identical results in Table \ref{tab:cf_results_nonrejectpoisson} in Appendix \ref{sec:cf_results_nonrejectpoisson} for the subset of clinics for which we cannot reject that the machine learning predictions are unbiased. For these, selection on unobservables into testing is least likely.} The first counterfactual provides physicians with the machine learning prediction of type $\tau_i$ for every patient and assumes that physicians use it without noise by setting $\sigma_{\xi_{j}}=0$. In this counterfactual, clinical diagnostic skill and the payoff function parameter are held fixed. We find that overall prescribing decreases by 25.4 percent (4,140 prescriptions) and overprescribing decreases by 44.5 percent (2,839 prescriptions). Interestingly, the number of treated bacterial infections is also reduced by 13.1 percent (1,302 prescriptions). Hence, the improved and more precise information on patient type shows how conservative physicians are on average. In the context of antibiotic prescribing in primary care overprescribing is the foremost concern because effects of antibiotic treatment cannot be reversed while decisions to delay prescribing can be corrected as soon as complete test results are available.

\begin{table}[t!]
\centering
\small
\begin{threeparttable}
\caption{Counterfactual policy outcomes}
\label{tab:cf_results}
\begin{tabular}{ @{} l  @{\extracolsep{2mm}} c @{\extracolsep{3mm}} c @{\extracolsep{3mm}} c @{} } \\[-1cm]
\toprule
			 							& 1. Provide ML-based $\tau_i$ 		& 2. Manipulate payoffs  					& 3. Redistribution 	\\[0.1em]
				\cline{2-2}\cline{3-3}\cline{4-4}\\[-1.5em]
			 							& Set $\sigma_{\xi_{j}}=0$				& $\beta^{\prime}_j = \min(\beta_j + \kappa,1)$	& $\Delta \delta y \overset{!}{=} 0$		\\
\midrule
Overall prescribing, $\Delta \delta$ in percent		&\multirow{2}{*}[9pt]{ -25.4 [-26.1, -23.1]} 	&\multirow{2}{*}[9pt]{-25.4 [-26.4, -24.5]} 		&\multirow{2}{*}[9pt]{-10.0 [-11.4, -9.5]} 	\\[-0.5em]
$N_\delta = 16,334$	 						&				 				&						 			&						 		\\
Treated bacterial cases, $\Delta \delta y$ in percent	&\multirow{2}{*}[9pt]{-13.1 [-14.6, -10.6]} 	&\multirow{2}{*}[9pt]{-20.4 [-22.2, -18.7]} 		&  \multirow{2}{*}[9pt]{$\ $0 [0, 0]} 		\\[-0.5em]
$N_{\delta y} = 9,957$	 					&				 				&									&   								\\
Overprescribing, $\Delta \delta \neg y$ in percent	&\multirow{2}{*}[9pt]{-44.5 [-45.3, -41.3]} 	&\multirow{2}{*}[9pt]{-33.1 [-35.0, -31.4]} 		& \multirow{2}{*}[9pt]{-25.2 [-28.8, -24.6]} \\[-0.5em]
$N_{\delta \neg y} = 6,377$	 				&				 				&									&						 		\\
\midrule
Change in payoffs, $W_j(\delta_j)$ in percent		& 20.1	&	5.8	& 12.1 \\[0.2em]
\bottomrule
\end{tabular}
\vspace{-0.2cm}
\footnotesize Notes: This table reports changes to the status quo in percent across 194 clinics. The observed absolute numbers are reported in the left column. In counterfactual two, we set $\kappa=0.12$ to obtain the same $\Delta \delta$ as in counterfactual one.
\end{threeparttable}
\end{table}

In counterfactual two, we manipulate the physicians' payoff functions but keep physician skill regarding patient type and clinical diagnostics unchanged. In particular, we increase the payoff parameter $beta_{j}$ by a constant $\kappa$ such that the overall reduction in prescribing is equivalent to the counterfactual reduction achieved by providing the machine learning predictions to physicians. More precisely, we define the counterfactual payoff parameter as $\beta^{\prime}_j = \min(\beta_j + \kappa,1)$, where setting $\kappa=0.12$ attains the overall reduction in prescribing of 25.4 percent. Such an intervention can be interpreted as a nudge or an antibiotic tax that shifts the relative weights on the social cost of increasing antibiotic resistance and an individual patients' sickness cost of foregone antibiotic treatment. Although overall reduction in prescribing is the same as in counterfactual one, overprescribing is reduced significantly less by 33.1 percent (2,110 prescriptions), compared to 44.5 percent in the first counterfactual. Notably, manipulating the payoff function weights without improving diagnostic information induces adverse effects as reflected in a large decrease in treated bacterial infections by 20.4 percent (2,207 prescriptions).

These results illustrate the usefulness of separating the prediction and decision step in the structural model. The effects of interventions attempting to incentivize behavior according to social objectives can be considered independently from interventions aimed at purely improving diagnostic information. This is in contrast to situations studied by Cowgill and Stevenson (2020) in which algorithm outputs are manipulated to communicate not only predictions but also social objectives. They argue that such manipulations can lead to refusal by human experts to use predictions. The framework we consider allows for interventions in which the two aims, providing machine learning predictions to experts and incentivizing social behavior, can be implemented and evaluated as complements.

In counterfactual three, we evaluate a prescription rule that corresponds to the procedure used in the prior literature evaluating machine learning predictions (Bayati et al. 2014; Chalfin et al. 2016; Kleinberg et al. 2018; Ribers and Ullrich 2019; Yelin et al. 2019; Hastings, Howison, and Inman 2020). The rule relies on the assumption that human discretion can be overruled or that decision makers adhere perfectly to prediction-based prescription rules. In this policy, prescriptions for patients with low predicted risk are delayed until test results are available. All patients with high predicted risk receive prescriptions before test results arrive. As the payoff function parameter $\beta_j$ is typically unknown in this stream of literature, it is convenient to focus on a corner solution that maximally reduces antibiotic use while enforcing $\Delta \delta y = 0$. This guarantees a welfare increase to a social planner for all $\beta^{S} \in [0,1]$. This counterfactual policy reduces overall prescribing by 10.0 percent (1,634 prescriptions) and overprescribing by 25.2 percent (1,607 prescriptions) while, by construction, the change in the number of prescriptions to bacterial infections is zero. We note the significant differences compared to the outcomes of counterfactual one where physicians incorporate patient type information provided by the machine learning predictions. 

Table \ref{tab:cf_results} also reports the mean change of physician payoffs in Equation (\ref{eqn_policy_payoff_function}) for each policy, defined as 
\begin{equation}
	W_{j}(\delta_j) 
		= \frac{ \Pi_{j}(\delta_j) - \Pi_{j}(\delta_j^0) }{ \bar{\Pi}_{j} - \Pi_{j}(\delta_j^0) }
\end{equation}
where $\bar{\Pi}_j = - \hat{\beta}_j \sum_{i\in \mathcal{I}{j}} y_i$ is the first best outcome realized if and only if the physician gives prescriptions to patients with a bacterial infection and $\Pi_j(\delta_j) = \sum_{i\in \mathcal{I}{j}} \pi(d_{ij},y_{i} \, ; \, \hat{\beta}_{j}) $ is the physician's payoff for the set of decisions $d_{ij} \in \delta_j$. Payoff gains are largest for the counterfactual policy which provides patient type information and smallest for the policy manipulating payoff weights on the antibiotic resistance externality. This result is expected because increasing the weight on the cost of prescribing resembles a tax reducing utility. The gains from the redistribution policy, which constrains physician discretion, lie in between. The results based on counterfactual policies one and two provide a more realistic \emph{ex ante} evaluation of the potential gains due to machine learning predictions because they do not rely on the assumption that physicians adhere perfectly to a prediction-based decision rule. Complete replacement of human decision makers is often not desirable or realistic, making it difficult to draw conclusions from results relying on such an assumption. The differences in payoff gains reflect that implementation of a redistribution policy may conflict with physician and patient preferences.

Finally, taking the perspective of a social planner, we evaluate welfare effects to provide a ranking of counterfactual policies based not only on reported counts of outcomes but on gains in payoffs given social preferences. We calculate the welfare effects for the three counterfactual policies over the continuum of potential social planner preference parameter values $\beta^{S} \in [0,1]$ as
\begin{equation}
	W(\delta, \beta^{S}) 
		= \frac{ \Pi(\delta, \beta^S) - \Pi(\delta^0, \beta^S) }{ \bar{\Pi} - \Pi(\delta^0, \beta^S) }
\end{equation}
where $\bar{\Pi} = - \beta^{S} \sum_{i\in \mathcal{I}} y_i$ is the first best aggregate outcome over the full set of patients, $\mathcal{I}$, realized if and only if prescriptions are given to patients with a bacterial infection and $\Pi(\delta, \beta^S) = \sum_{i\in \mathcal{I}} \pi(d_{ij(i)},y_{i} \, ; \, \beta^{S}) $ is the aggregated payoff function for the set of decisions $d_{ij(i)} \in \delta$.

\begin{figure*}[h!]
	\centering
	\includegraphics[height=10.5cm]{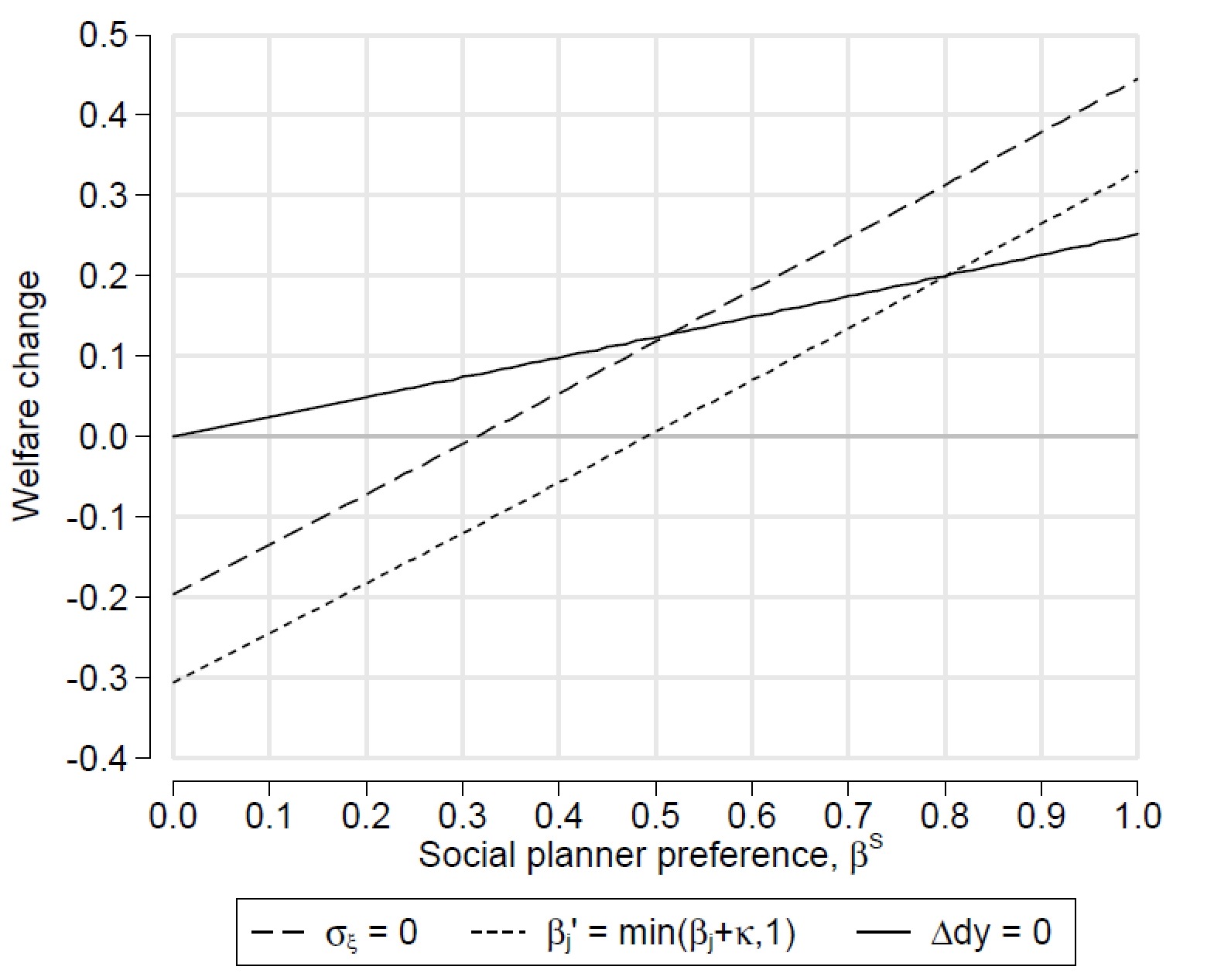}
    	\caption{Counterfactual welfare effects for social planner preferences $\beta^S\in[0,1]$.}
	 \label{fig:welfare}
\end{figure*}
Figure \ref{fig:welfare} shows $W(\delta, \beta^{S})$ over the full support of $\beta^{S}$, revealing that the best policy depends on the social planner's weight on the antibiotic resistance externality. The third counterfactual is based on the constraint that prescribing to bacterial infections cannot be reduced, which guarantees that positive payoff gains are always achieved. This is particularly important if preferences are unknown. However, with known preferences, the counterfactuals which give physicians full discretion and make use of their expertise dominate the corner solution. This is intuitive because, if society places a sufficiently large weight on the externality, then it is worthwhile even to reduce prescriptions to some bacterial infections in exchange for a reduction of overprescribing. The preference parameter above which counterfactuals one and two achieve higher gains than counterfactual three, lies just below the mean estimated physician-level $\beta_j$ of $0.56$. Therefore, if the social planner's weight $\beta^S$ is at least as large as the average physician's $\beta_j$, providing better diagnostic information to physicians yields the highest social benefits.

\section{Conclusion}

We show how policies enabled by machine learning predictions can be evaluated when humans hold decision-relevant information. It is typically difficult to determine whether such information is complementary to or can be substituted by machine learning predictions. If such information, or the skill required to obtain it, is difficult to measure and varying across decision makers, assessing the added value of machine learning predictions \emph{ex ante} is challenging. Field trials may be designed to provide reliable assessments but are often difficult to implement for ethical, legal, or practical reasons. It is therefore important to develop model-based tools to evaluate potential implementations \emph{ex ante}. Promising evaluations may help convince stake holders that field trials are worthwhile to implement. 

The setting we consider for this analysis, antibiotic prescribing for suspected urinary tract infections, resembles many situations in primary care provision but also expert decision problems more generally. While we consider and exploit some idiosyncrasies of the primary care setting, our analysis provides a more generally applicable framework for the evaluation of machine learning in data-rich environments. Whether and how our analysis can be helpful in alternative settings depends on measurement of the target outcome and on the availability of data to allow unbiased predictions at the level of decision makers. We provide evidence supporting the assumption that, combined with the ability to observe the full set of outcomes, observable patient information is sufficiently rich to identify skill and payoffs in our setting. Unbiasedness may not only be threatened by insufficient prediction quality but also by potential selection problems when decisions impact measured outcomes. In such settings, alternative identification strategies, for example based on random assignment assumptions as proposed by Kleinberg et al. (2018) and Chan, Gentzkow, and Yu (2020), are necessary in addition to quality machine learning predictions.

Several important avenues for further research specific to the context of antibiotic prescribing remain. It would be worthwhile to attempt encoding further clinical information, for example, from electronic health records, such as reported symptoms and results from in-clinic diagnostics to further improve machine learning predictions. We also omit an important dimension of antibiotic prescribing, the choice of molecule. It remains an open question for future research to what extent prediction of bacterial species and molecule-specific resistances can further inform prescription choice. Further research is needed to better understand experts' potential behavioral reactions to the introduction of prediction tools. Results from such studies may provide insights on how to optimally communicate machine learning predictions, to what extent to explain prediction outcomes, and potential effects on decision makers' incentives to acquire information and expertise.

While we consider one specific type of medical diagnostic and treatment problem, our results indicate the potential of using machine learning predictions in many relevant health care situations. Data availability and the quality of prediction algorithms are improving at a rapid pace in and beyond health care. The rate at which such technologies will be more broadly adopted and productively exploited, however, will depend on the kind and quality of human expertise it can complement. If our findings are suggestive more generally, human experts are far from being replaced. Instead, investment in human capital is likely one key to help fulfill the promise of welfare-improving technological progress.


\clearpage
\begin{appendices}

\section{Physician heterogeneity}
\label{app:gp_aucreg}

\begin{table}[h]
\centering
\small
\begin{threeparttable}
\caption{Physician decisions as predictors and clinic characteristics}
\label{tab:gp_aucreg}
\begin{tabular}{ @{} l @{\extracolsep{10mm}} r @{\extracolsep{3mm}} l @{} }
\toprule
Outcome: $\Delta$ AUC from including	 & \multicolumn{2}{c}{linear regression coefficients}  \\
treatment decisions as predictor				 & 	 \\
\midrule
Number of clinic's unique patients per physician 	& 0.007& [-0.015, 0.028] \\
Number of laboratory results per unique patient  	& 0.007& [-0.005, 0.018]  \\
Mean number of physicians  					&-0.002& [-0.017, 0.014]  \\
Mean age of physicians  						&-0.023& [-0.074, 0.027]  \\
Share of female physicians  					& 0.007& [-0.002, 0.015]  \\
\midrule
R$^2$				  					& \multicolumn{2}{c}{0.05}  \\
Observations			  					& \multicolumn{2}{c}{117}  \\
\bottomrule
\end{tabular}
\vspace{-0.2cm}
\footnotesize Notes: Heteroskedasticity-robust 95\% confidence intervals in brackets. 
\end{threeparttable}
\end{table}

\clearpage
\section{Signals on patient type and sickness state}
\label{app:model_signals}

\begin{figure*}[h!]
	\centering
	\includegraphics[height=10cm]{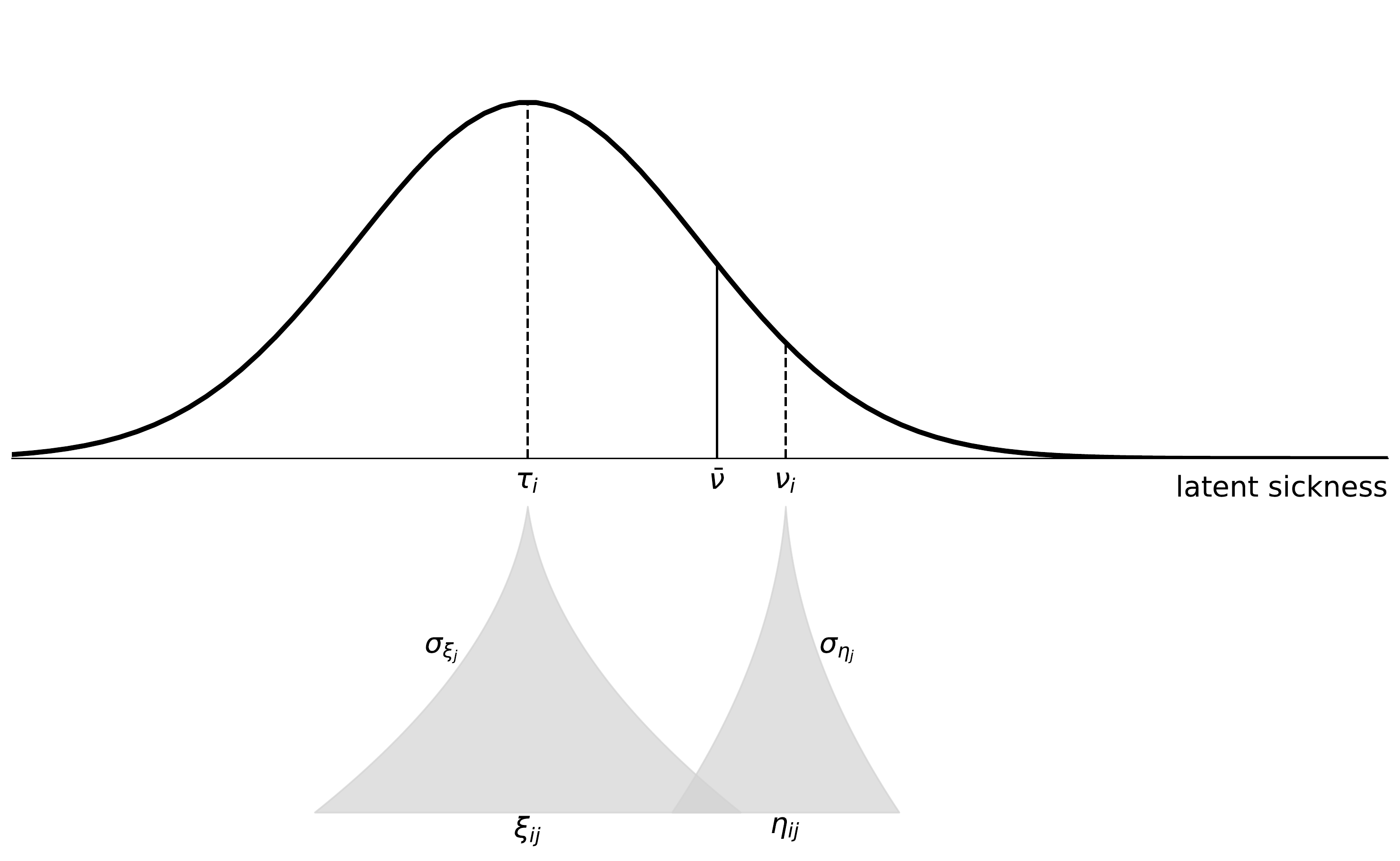}
    	\caption{The distribution of $\nu_i$ and the signals $\xi_{ij}$ and $\eta_{ij}$ for an example patient, who has a bacterial infection ($\nu_i>\bar{\nu}$).}
	 \label{fig:model_signals}
\end{figure*}

\clearpage
\section{Simulated choice probabilities} \label{app:sim_smooth}
Simulation procedure for patient $i$ and physician $j$ (index $i$ and $j$ suppressed below):
\begin{enumerate}
\item Given $m(x)$ and $y$, draw simulated sickness realizations
\begin{equation*}
	\nu^{r} \sim \mathcal{N}(m(x), 1 \mid y = \mathbbm{1}_{[\nu^{r} > \bar{\nu}]}).
\end{equation*}
and signals
\begin{align*}
	\xi^{r} &\sim \mathcal{N}(m(x), \sigma_{\xi}^{2}) \\
	\eta^{r} &\sim \mathcal{N}(\nu^{r}, \sigma_{\eta}^{2})
\end{align*}
for physician parameters $\sigma_{\xi}$ and $\sigma_{\eta}$.

\item Compute the expected payoff:
\begin{equation*}
	EU_{d}^{r} = \E \{ \pi \mid d, \xi^{r}, \eta^{r} \} =
	 \begin{cases} \\[-1.5cm]
		 -\alpha  \overbracket{ \left(1 - \Phi \left( \frac{\bar{\nu}-\mu^{r}}{\sigma^{r}} \right) \right) }^{\Pr\{ y=1 \mid \: \xi^{r}, \eta^{r} \}}  
		 & \text{if } d = 0 \\
		 -\beta
		 & \text{if } d = 1
	 \end{cases}					
\end{equation*}
where $\alpha$ and $\beta$ are physician parameters and $\mu^{r}$ and $\sigma^{r}$ are functions of $\xi^{r}, \eta^{r}, \sigma_{\xi},\sigma_{\eta}$ and $\sigma_{\eta}$ as stated in equation~\ref{eq:posterior_m_s}. Note that either $\alpha$ or $\beta$ must be normalized. We set $\alpha=1$.

\item Computing the simulated choice probabilities
\begin{equation*}
	\hat{P}_{0} = \frac{1}{R} \sum_{r=1}^{R} \mathbbm{1}_{[EU_{0}^{r}>EU_{1}^{r}]}
	\qquad \text{and} \qquad
	\hat{P}_{1} = \frac{1}{R} \sum_{r=1}^{R} \mathbbm{1}_{[EU_{1}^{r} \geq EU_{0}^{r}]}
\end{equation*}
yields step-functions of $\sigma_{\xi}$, $\sigma_{\eta}$ and $\beta$, which are difficult to minimize numerically in maximum likelihood estimation. Therefore, we compute choice probabilities using the Logit-smoothed Accept-Reject simulator with smoothing parameter $\lambda=0.01$. We insert the utilities into the logit formula:
\begin{equation*}
	S_{0}^{r} =
			\frac{	e^{\nicefrac{EU_{0}^{r}}{\lambda}}}
			{e^{\nicefrac{EU_{0}^{r}}{\lambda}}+e^{\nicefrac{EU_{1}^{r}}{\lambda}}}
			= \frac{1}{1 + e^{\frac{1}{\lambda} (1-\Phi \left( \frac{\bar{\nu}-\mu^{r}}{\sigma^{r}} \right)-\beta)} }
\end{equation*}
and
\begin{equation*}
	S_{1}^{r} = 
			\frac{	e^{\nicefrac{EU_{1}^{r}}{\lambda}}}
			{e^{\nicefrac{EU_{0}^{r}}{\lambda}}+e^{\nicefrac{EU_{1}^{r}}{\lambda}}}
			= \frac{1}{1 + e^{\frac{1}{\lambda} (\Phi \left( \frac{\bar{\nu}-\mu^{r}}{\sigma^{r}} \right)-1+\beta)} }.
\end{equation*}
\item Repeat steps 1-3 $R$ times, so that $r$ takes the values 1 through $R$.
\item We obtain the simulated choice probabilities by averaging over simulations:
\begin{equation*}
	\hat{P}_{0} = \frac{1}{R} \sum_{r=1}^{R} S_{0}^{r}
	\qquad \text{and} \qquad
	\hat{P}_{1} = 1- \hat{P}_{0} =  \frac{1}{R} \sum_{r=1}^{R} \left( 1-S_{0}^{r} \right) =  \frac{1}{R} \sum_{r=1}^{R} S_{1}^{r}
\end{equation*}
\end{enumerate}

\clearpage
\section{Analytical gradients}
\label{agradients}

The objective function can be written
\begin{equation*}
	LLH(\sigma_{\xi_{j}},\sigma_{\eta_{j}},\beta_{j} \mid \xi_{ij}, \eta_{ij}, d_{ij}) 
		= \sum_{i\in \mathcal{I}_{j}} \log \left( \mathcal{L}_{ij} \left( d_{j} \mid \Theta_{j}, y_{j}, m(x_{i}) \right) \right)
		= \sum_{i\in \mathcal{I}_{j}} d_{ij} \log ( \hat{P}_{ij1} ) + (1-d_{ij}) \log (1- \hat{P}_{ij1} )
\end{equation*}
The derivatives of the LLH function for a single datapoint, dropping index $ij$ for simplicity, are given by:
\begin{align*}
	\frac{\partial LLH}{\partial \beta} 
			&= \left( \frac{d}{\hat{P}_{1}} \right) \frac{\partial \hat{P}_{1}}{\partial \beta}
			+ \left(\frac{1-d}{1-\hat{P}_{1}} \right) \left( - \frac{\partial \hat{P}_{1}}{\partial \beta} \right) \\
			& = \left( \frac{d}{\hat{P}_{1}} - \frac{1-d}{1-\hat{P}_{1}} \right) \frac{\partial \hat{P}_{1}}{\partial \beta} \\
			& = \left( \frac{d}{\hat{P}_{1}} - \frac{1-d}{1-\hat{P}_{1}} \right) \frac{1}{R} \sum_{r=1}^{R} S_{1}^{r} (1-S_{1}^{r}) \frac{\partial \frac{1}{\lambda}(EU_{0}^{r}-EU_{1}^{r})}{\partial \beta} \\
			& = \left( \frac{d}{\hat{P}_{1}} - \frac{1-d}{1-\hat{P}_{1}} \right) \frac{1}{R} \sum_{r=1}^{R} S_{1}^{r} (1-S_{1}^{r}) \frac{\partial \frac{1}{\lambda} (\Phi \left( - \frac{\mu^{r}}{\sigma^{r}} \right)-1+\beta)}{\partial \beta} \\
			& = \left( \frac{d}{\hat{P}_{1}} - \frac{1-d}{1-\hat{P}_{1}} \right) \frac{1}{R} \sum_{r=1}^{R} S_{1}^{r} (1-S_{1}^{r}) \frac{1}{\lambda}
\intertext{and}
	\frac{\partial LLH}{\partial \sigma_{\xi}}
		&= \left( \frac{d}{\hat{P}_{1}} - \frac{1-d}{1-\hat{P}_{1}} \right) \frac{1}{R} \sum_{r=1}^{R} S_{1}^{r} (1-S_{1}^{r}) \frac{1}{\lambda} \frac{\partial \Phi \left( -\frac{\mu^{r}}{\sigma^{r}} \right)}{\partial \sigma_{\xi}} \\
		&= \left( \frac{d}{\hat{P}_{1}} - \frac{1-d}{1-\hat{P}_{1}} \right) \frac{1}{R} \sum_{r=1}^{R} S_{1}^{r} (1-S_{1}^{r}) \frac{1}{\lambda} \phi \left( -\frac{\mu^{r}}{\sigma^{r}} , 0, 1 \right) \frac{\partial \left( -\frac{\mu^{r}}{\sigma^{r}} \right) }{\partial \sigma_{\xi}}
\intertext{with}
	\frac{\partial \left( -\frac{\mu^{r}}{\sigma^{r}} \right) }{\partial \sigma_{\xi}}
	&= - \frac{\sigma_{\eta} \left(\eta_{N} \sigma_{\eta} \sigma_{\xi}^{3} + \eta_{N} \sigma_{\eta} \sigma_{\xi} - \sigma_{\eta}^{2} \sigma_{\xi} \tau + \sigma_{\eta}^{2} \xi_{N} - \sigma_{\xi}^{4} \xi_{N} - 2 \sigma_{\xi}^{3} \tau + \sigma_{\xi}^{3} \nu - 2 \sigma_{\xi} \tau + \sigma_{\xi} \nu + \xi_{N}\right)}{\left(\sigma_{\eta}^{2} \sigma_{\xi}^{2} + \sigma_{\eta}^{2} + \sigma_{\xi}^{4} + 2 \sigma_{\xi}^{2} + 1\right)^{\frac{3}{2}}}
\intertext{and}
	\frac{\partial LLH}{\partial \sigma_{\eta}} &= \left( \frac{d}{\hat{P}_{1}} - \frac{1-d}{1-\hat{P}_{1}} \right) \frac{1}{R} \sum_{r=1}^{R} S_{1}^{r} (1-S_{1}^{r}) \frac{1}{\lambda} \phi \left( -\frac{\mu^{r}}{\sigma^{r}} , 0, 1 \right) \frac{\partial \left( -\frac{\mu^{r}}{\sigma^{r}} \right) }{\partial \sigma_{\eta}}
\intertext{with}
	\frac{\partial \left( -\frac{\mu^{r}}{\sigma^{r}} \right) }{\partial \sigma_{\eta}}
	&= \frac{\eta_{N} \sigma_{\eta}^{3} \sigma_{\xi}^{2} + \eta_{N} \sigma_{\eta}^{3} - \sigma_{\eta}^{2} \sigma_{\xi}^{3} \xi_{N} - \sigma_{\eta}^{2} \sigma_{\xi}^{2} \tau + 2 \sigma_{\eta}^{2} \sigma_{\xi}^{2} \nu - \sigma_{\eta}^{2} \sigma_{\xi} \xi_{N} - \sigma_{\eta}^{2} \tau + 2 \sigma_{\eta}^{2} \nu + \sigma_{\xi}^{4} \nu + 2 \sigma_{\xi}^{2} \nu + \nu}{\sigma_{\eta}^{2} \sqrt{\sigma_{\xi}^{2} + 1} \left(\sigma_{\eta}^{2} + \sigma_{\xi}^{2} + 1\right)^{\frac{3}{2}}}
\end{align*}

\clearpage
\section{Parameter estimates} \label{sec:est_struct_par}

\begin{figure*}[h!]
	\centering
	\includegraphics[height=8cm]{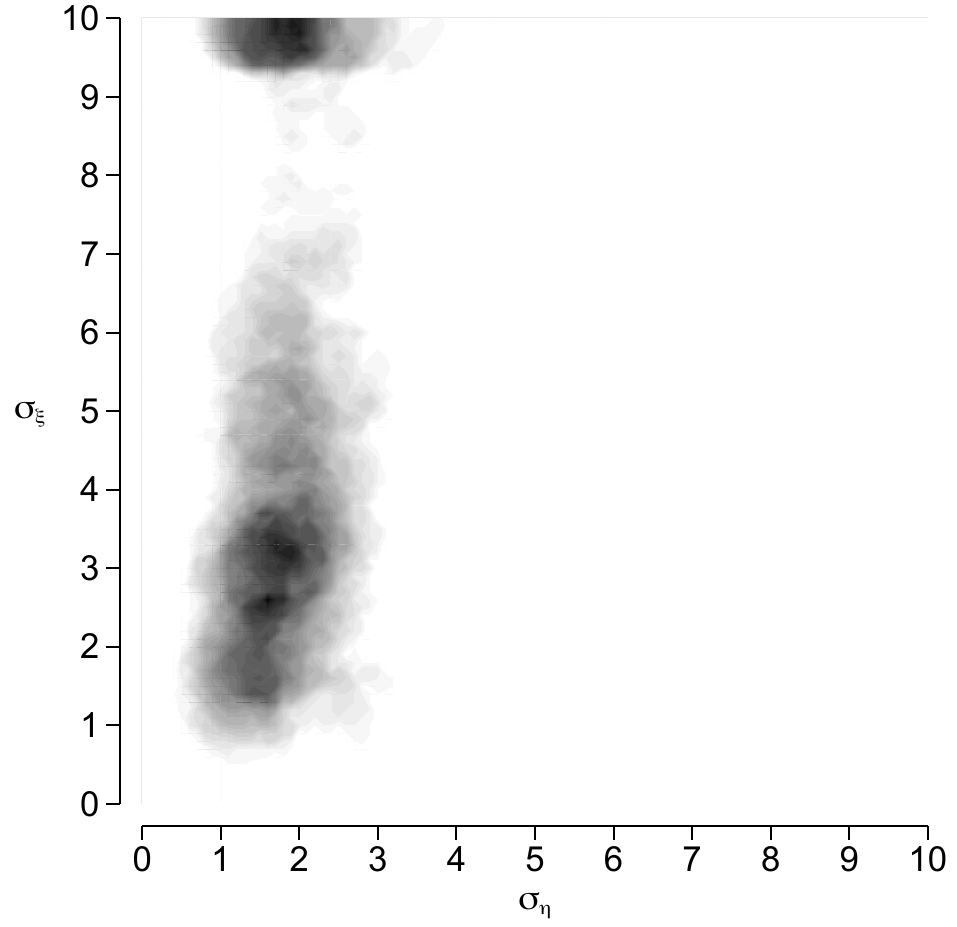}
    	\caption{Heatmap of physician-level estimates for $\sigma_{\xi}$ and $\sigma_{\eta}$. To ensure anonymity, the figure shows a heatmap covering only areas of five physicians or more.}
	 \label{fig:est_struct_par_xieta}
\end{figure*}

\begin{figure*}[h!]
	\centering
	\includegraphics[height=8cm]{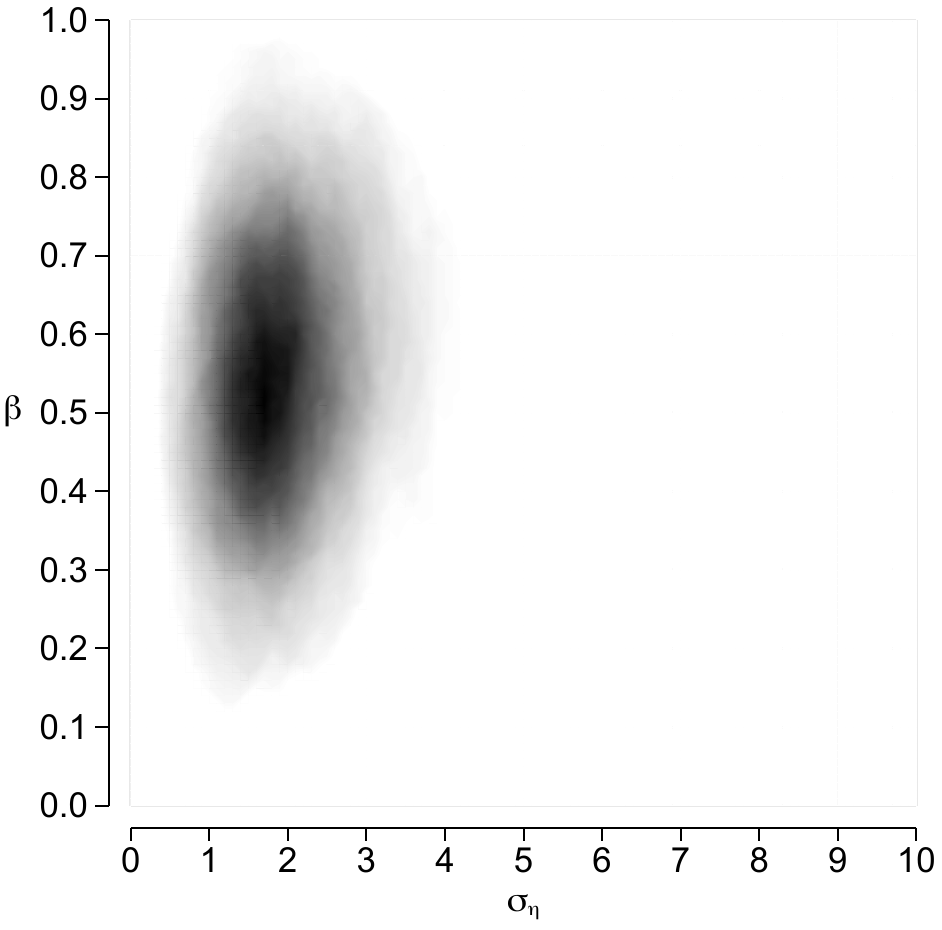}
    	\caption{Heatmap of physician-level estimates for $b$ and $\sigma_{\eta}$. To ensure anonymity, the figure shows a heatmap covering only areas of five physicians or more.}
	 \label{fig:est_struct_par_beta}
\end{figure*}

\begin{figure*}[h!]
	\centering
	\includegraphics[height=8cm]{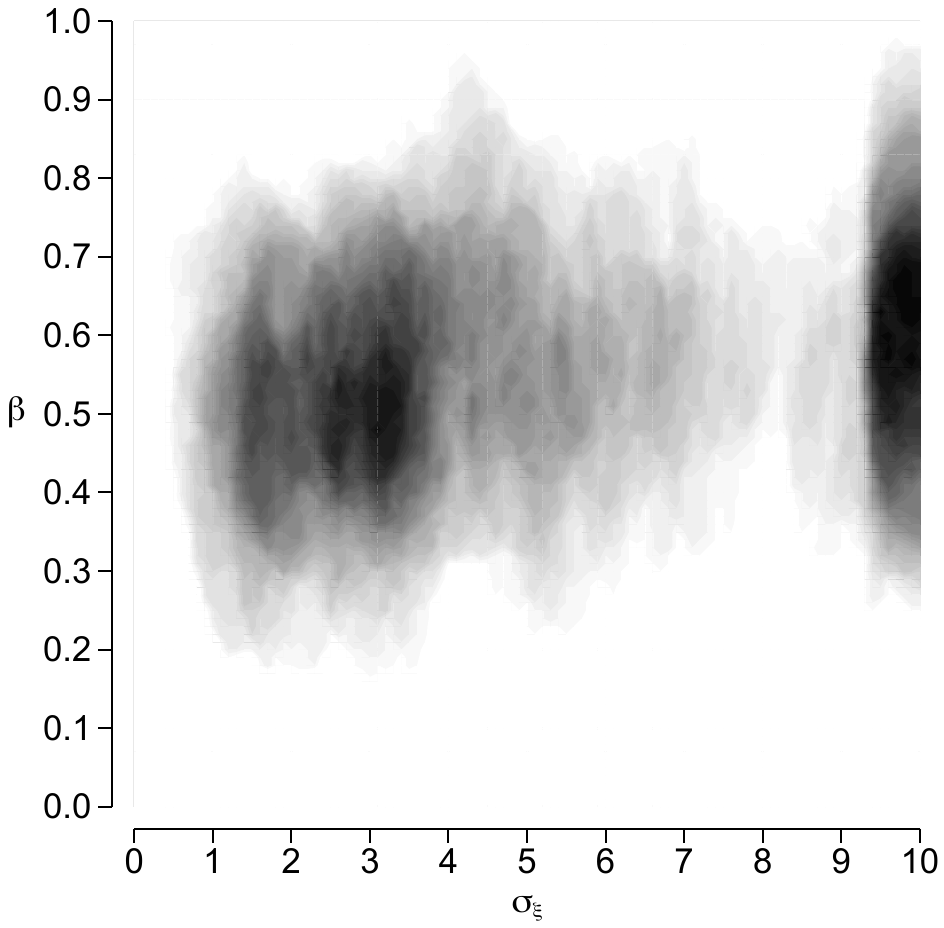}
    	\caption{Heatmap of physician-level estimates for $b$ and $\sigma_{\xi}$. To ensure anonymity, the figure shows a heatmap covering only areas of five physicians or more.}
	 \label{fig:est_struct_par_bxi}
\end{figure*}

\clearpage
\section{Variance of parameter estimates} \label{sec:bs_struct_par}

\begin{figure*}[h!]
	\centering
	\includegraphics[height=8cm]{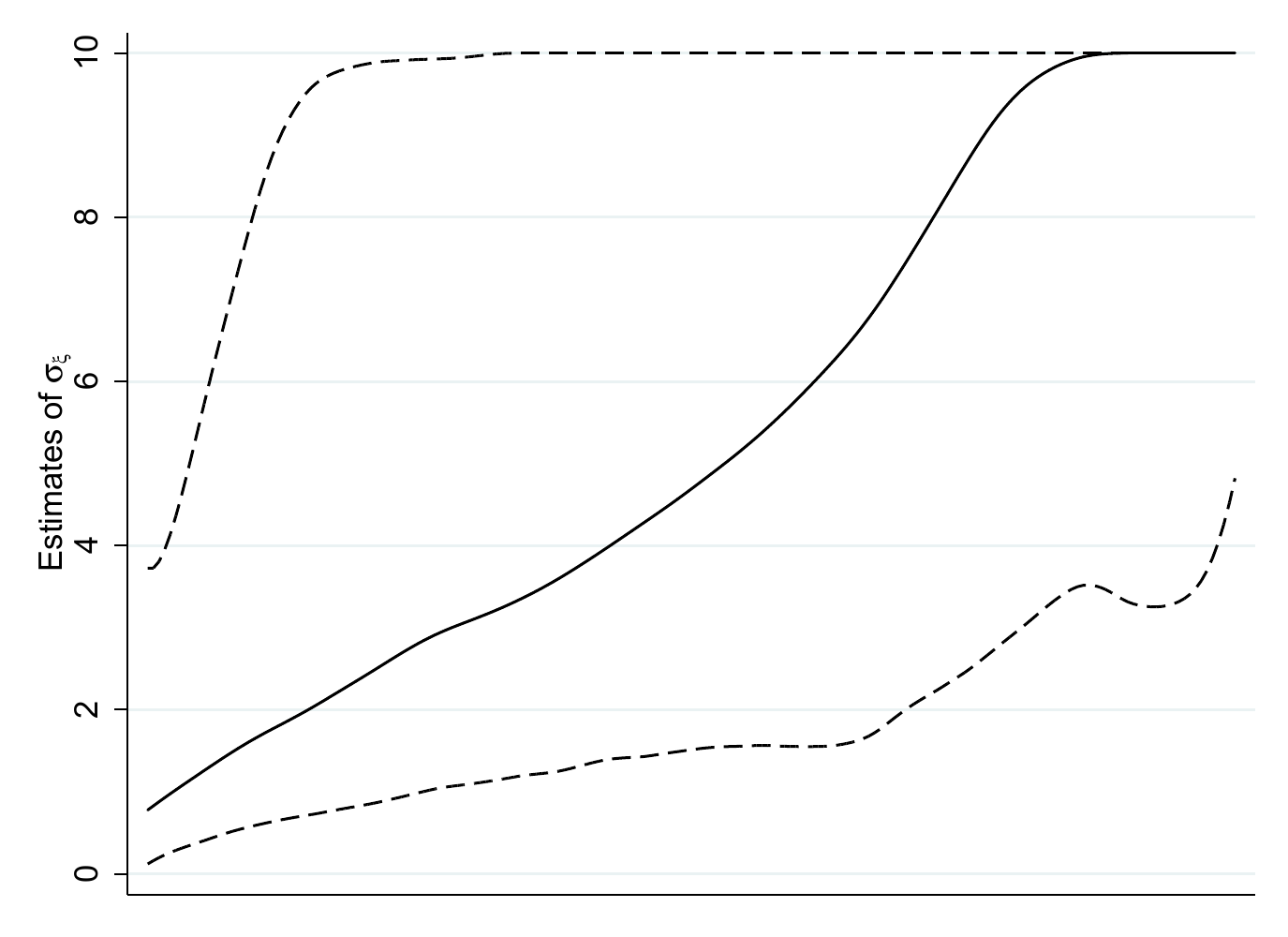}
    	\caption{Physician-level estimates and bootstrapped 95\% confidence intervals for $\sigma_{\xi}$. To ensure anonymity, the figure is LOWESS-smoothed with bandwidth 0.2.}
	 \label{fig:bs_struct_par_xi}
\end{figure*}

\begin{figure*}[h!]
	\centering
	\includegraphics[height=8cm]{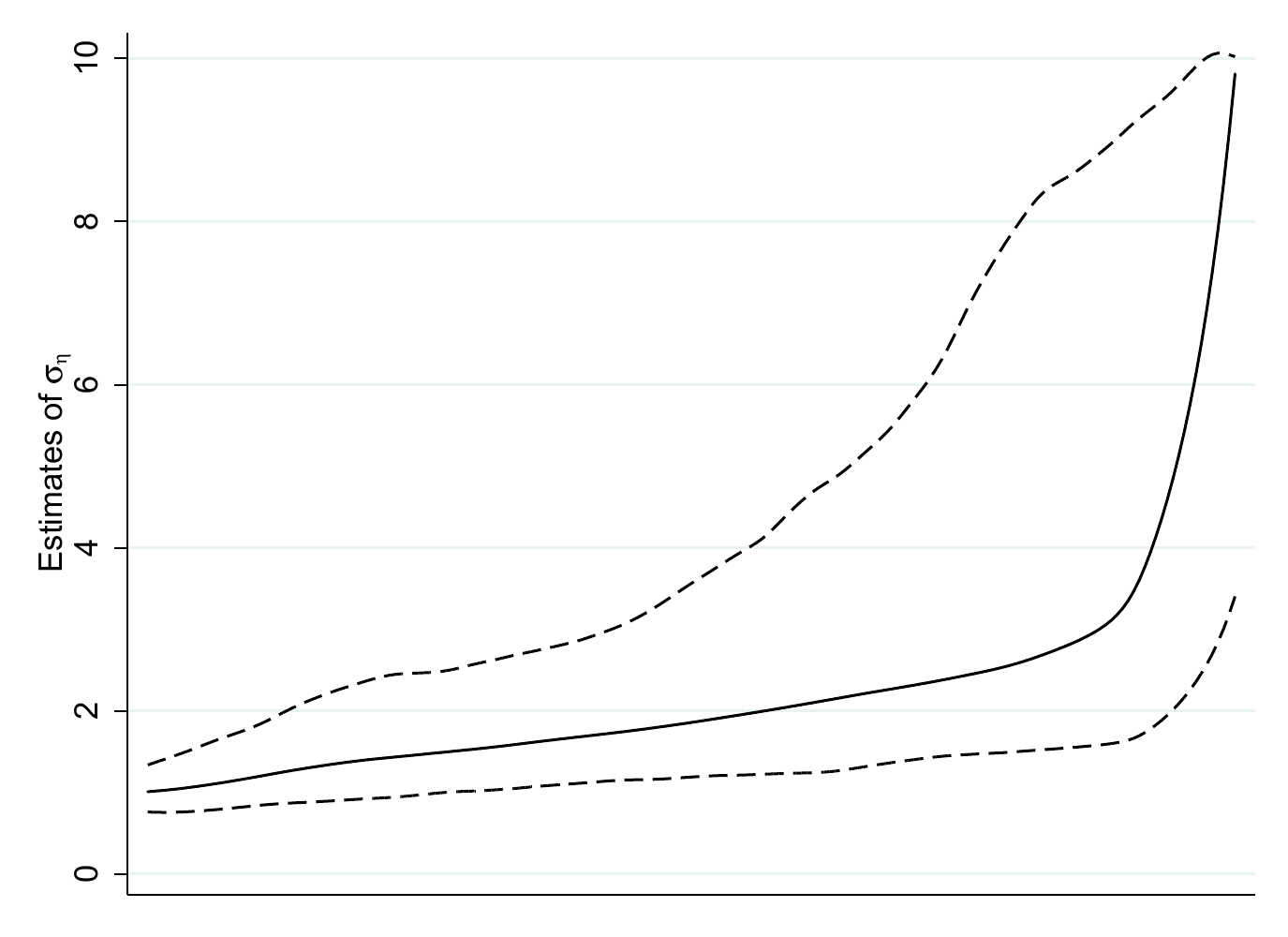}
    	\caption{Physician-level estimates and bootstrapped 95\% confidence intervals for $\sigma_{\eta}$. To ensure anonymity, the figure is LOWESS-smoothed with bandwidth 0.2.}
	 \label{fig:bs_struct_par_eta}
\end{figure*}

\begin{figure*}[h!]
	\centering
	\includegraphics[height=8cm]{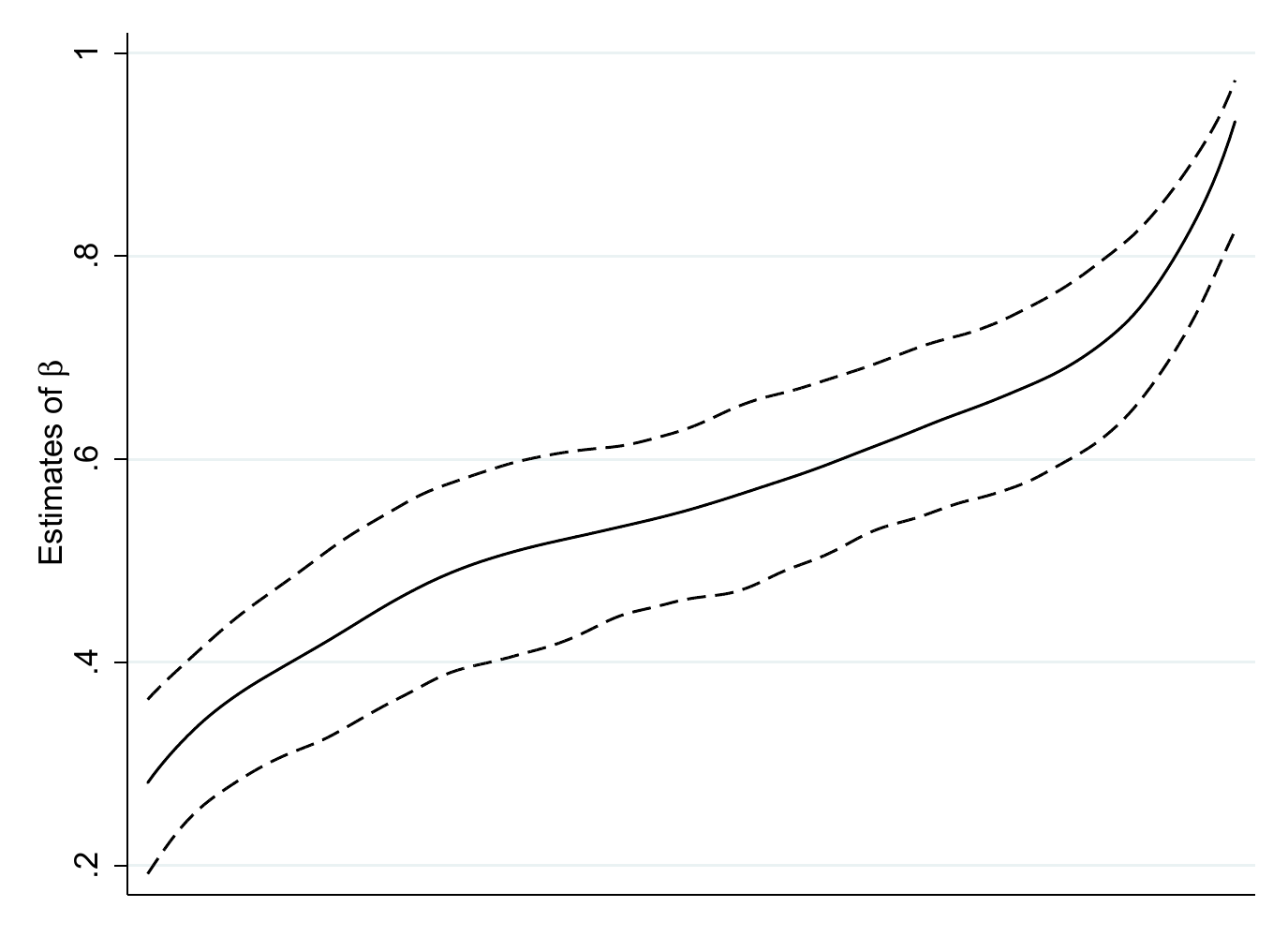}
    	\caption{Physician-level estimates and bootstrapped 95\% confidence intervals for $\beta$. To ensure anonymity, the figure is LOWESS-smoothed with bandwidth 0.2.}
	 \label{fig:bs_struct_par_b}
\end{figure*}

\clearpage
\section{Simulation of the decision-relevance of $\sigma_\xi$}
\label{sec:sim_rel_sdxi}

\begin{figure*}[h!]
	\centering
	\includegraphics[height=12cm]{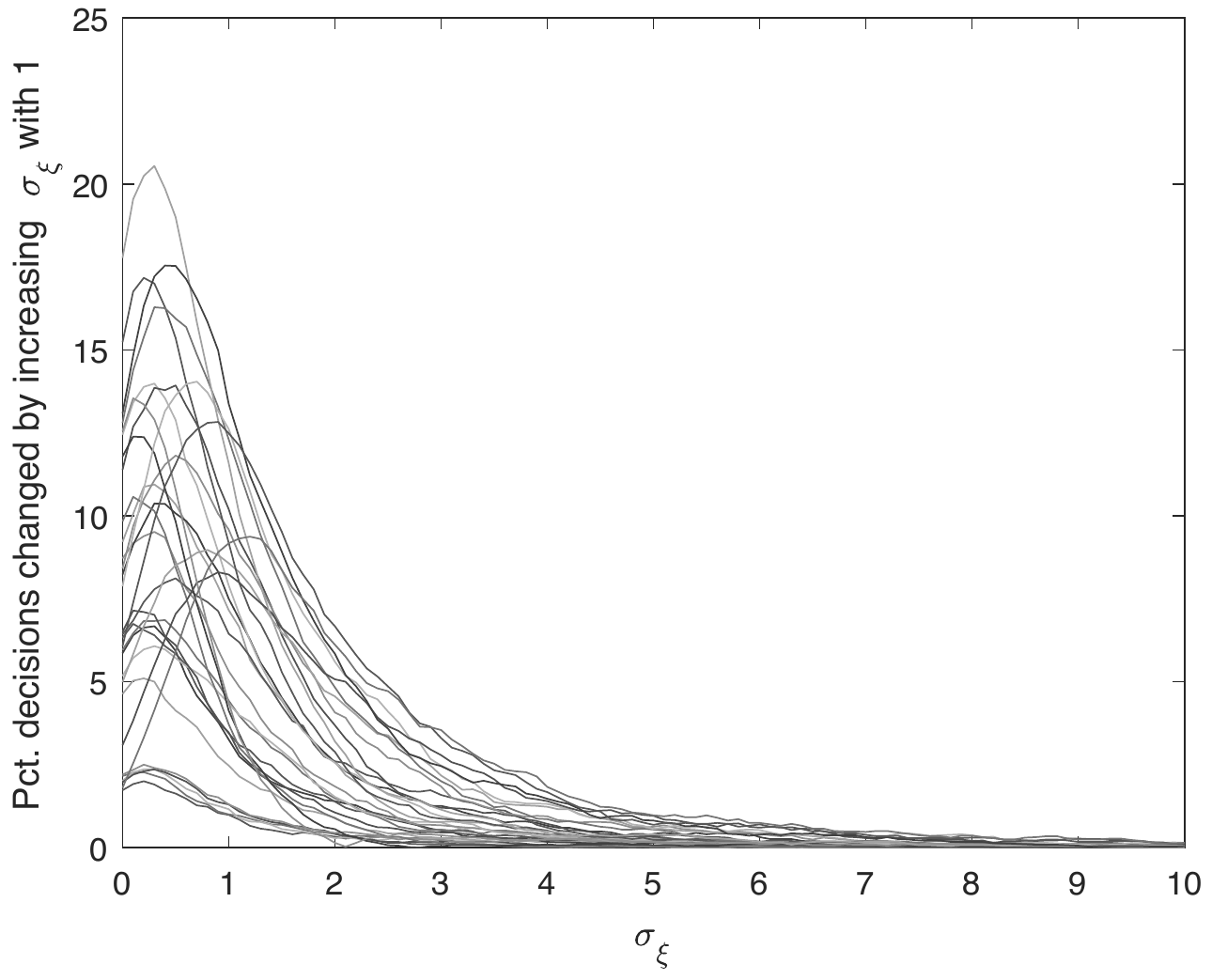}
    	\caption{Simulated changes in choice due to changes in $\sigma_\xi$, in percent per step size one.}
	 \label{fig:sim_rel_sdxi}
\end{figure*}

\clearpage
\section{Model fit}
\label{sec:est_struct_fit}

\begin{figure*}[h!]
	\centering
	\includegraphics[height=9cm]{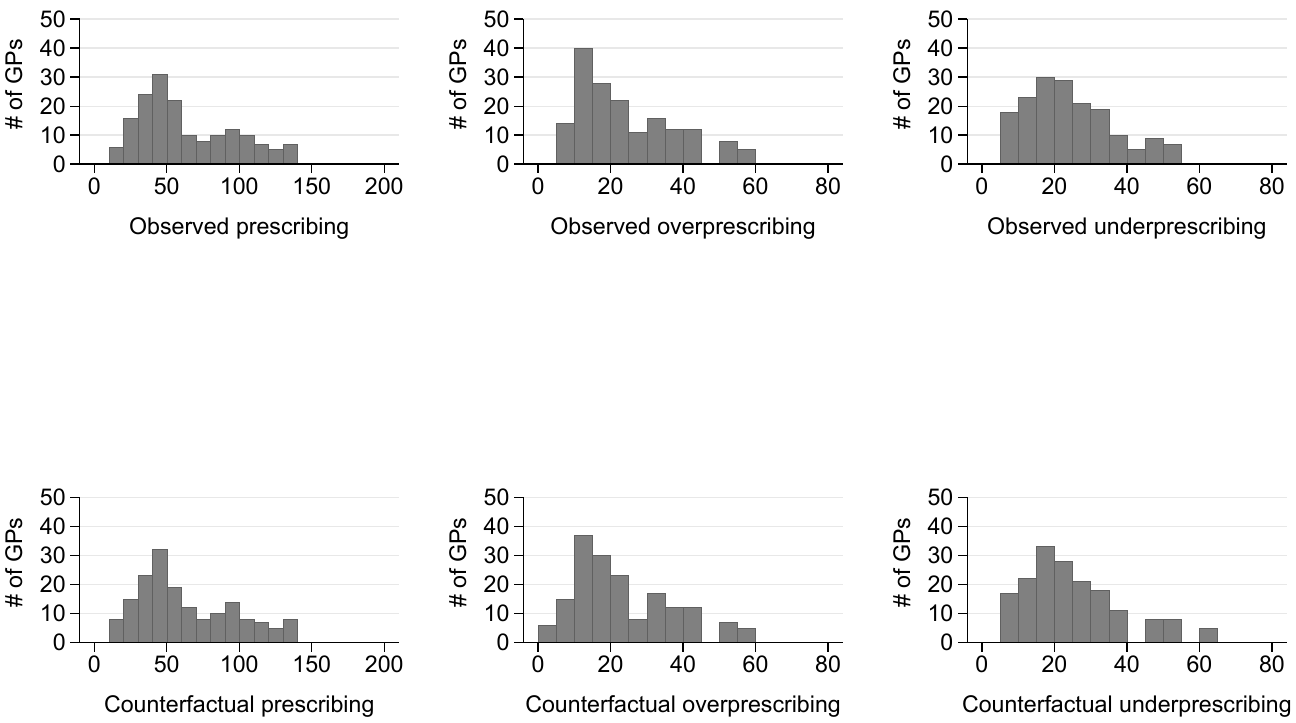}
    	\caption{Observed and simulated moments}
	 \label{fig:modelfit}
\end{figure*}

\clearpage
\section{Counterfactuals for reduced sample}
\label{sec:cf_results_nonrejectpoisson}

\begin{table}[h!]
\centering
\small
\begin{threeparttable}
\caption{Counterfactual policy outcomes, sample of physicians where unbiased prediction cannot be rejected}
\label{tab:cf_results_nonrejectpoisson}
\begin{tabular}{ @{} l  @{\extracolsep{2mm}} c @{\extracolsep{3mm}} c @{\extracolsep{3mm}} c @{} } \\[-1cm]
\toprule
			 						& 1. Provide ML-based $\tau_i$ & 2. Manipulate payoffs  	& 3. Redistribution 	\\[0.1em]
				\cline{2-2}\cline{3-3}\cline{4-4}\\[-1.5em]
			 						& Set $\sigma_{\xi_{j}}=0$				& $\beta^{\prime}_j = \min(\beta_j + \kappa,1)$  & $\Delta \delta y \overset{!}{=} 0$		\\
\midrule
Overall prescribing, $\Delta \delta$ in percent 	&\multirow{2}{*}[9pt]{ -25.5 [-26.3, -22.6]} 	&\multirow{2}{*}[9pt]{-25.5 [-26.2, -24.2]} 		&\multirow{2}{*}[9pt]{-9.92 [-11.2, -9.4]} 	\\[-0.5em]
$N_\delta = 13,484$	 					&				 				&						 			&						 		\\
Treated bacterial cases, $\Delta \delta y$ in percent	&\multirow{2}{*}[9pt]{-13.4 [-14.7, -10.5]} 	&\multirow{2}{*}[9pt]{-20.6 [-21.9, -18.7]} 		&  \multirow{2}{*}[9pt]{$\ $0 [0, 0]} 		\\[-0.5em]
$N_{\delta y} = 8,199$	 				&				 				&									&   								\\
Overprescribing, $\Delta \delta \neg y$ in percent		&\multirow{2}{*}[9pt]{-44.2 [-45.3, -40.4]} 	&\multirow{2}{*}[9pt]{-33.0 [-34.3, -30.5]} 		& \multirow{2}{*}[9pt]{-24.9 [-28.6, -24.2]} 	\\[-0.5em]
$N_{\delta \neg y} = 5,285$	 			&				 				&									&						 		\\
\midrule
Change in payoffs, $W_j(\delta_j)$ in percent		& 19.0	&	4.4	& 12.3 \\[0.2em]
\bottomrule
\end{tabular}
\vspace{-0.2cm}
\footnotesize Notes: This table reports changes to the status quo in percent across 160 clinics. The observed absolute numbers are reported in the left column. In counterfactual two, we set $\kappa=0.12$ to obtain the same $\Delta \delta$ as in counterfactual one.
\end{threeparttable}
\end{table}

\end{appendices}

\end{document}